\input harvmac\skip0=\baselineskip
\input epsf

\newcount\figno
\figno=0
\def\fig#1#2#3{
\par\begingroup\parindent=0pt\leftskip=1cm\rightskip=1cm\parindent=0pt
\baselineskip=11pt \global\advance\figno by 1 \midinsert
\epsfxsize=#3 \centerline{\epsfbox{#2}} \vskip 12pt {\bf Fig.\
\the\figno: } #1\par
\endinsert\endgroup\par
}
\def\figlabel#1{\xdef#1{\the\figno}}
\def\encadremath#1{\vbox{\hrule\hbox{\vrule\kern8pt\vbox{\kern8pt
\hbox{$\displaystyle #1$}\kern8pt} \kern8pt\vrule}\hrule}}



\lref\osv{
  H.~Ooguri, A.~Strominger and C.~Vafa,
  ``Black hole attractors and the topological string,''
  Phys.\ Rev.\ D {\bf 70}, 106007 (2004)
  [arXiv:hep-th/0405146].
}

\lref\WittenKT{
  E.~Witten,
  ``Three-Dimensional Gravity Revisited,''
  arXiv:0706.3359 [hep-th].
}

\lref\Tuite{
  M.~P.~Tuite,
  ``Genus Two Meromorphic Conformal Field Theory,''
  arXiv:math.qa/9910136.
}

\lref\Igusa{
  J.-I.~Igusa,
  ``On Siegel Modular Forms of Genus Two,"
  Am.J.Math. {\bf 84} (1962) 175-200;
  ``Modular Forms and Projective Invariants,"
  Am.J.Math. {\bf 89} (1967) 817-855.
}

\lref\FLM{ I.~Frenkel, J.~Lepowsky and A.~Meurman, ``A Natural
Representation of the Fischer-Griess Monster with the Modular
Function J as Character,'' Proc.Natl.Acad.Sci.USA {\bf 81} (1984)
3256-3260;
  I.~Frenkel, J.~Lepowsky and A.~Meurman,
  ``Vertex Operator Algebras and the Monster,''
{\it  Boston, USA: Academic (1988) 508 P. (Pure and Applied
Mathematics, 134)} }

\lref\ZamolodchikovAE{
  A.~B.~Zamolodchikov, ``Conformal Scalar Field on the
  Hyperelliptic Curve and Critical Ashkin-Teller Multipoint
  Correlation Functions,'' Nucl.\ Phys.\  B {\bf 285}, 481 (1987).
}

\lref\KnizhnikXP{
  V.~G.~Knizhnik,
  ``Analytic Fields on Riemann Surfaces. 2,''
  Commun.\ Math.\ Phys.\  {\bf 112}, 567 (1987).
}

\lref\DixonQV{
  L.~J.~Dixon, D.~Friedan, E.~J.~Martinec and S.~H.~Shenker,
  ``The Conformal Field Theory Of Orbifolds,''
  Nucl.\ Phys.\  B {\bf 282}, 13 (1987).
}

\lref\HamidiVH{
  S.~Hamidi and C.~Vafa,
  ``Interactions on Orbifolds,''
  Nucl.\ Phys.\  B {\bf 279}, 465 (1987).
}

\lref\MaldacenaBW{
  J.~M.~Maldacena and A.~Strominger,
  ``AdS(3) black holes and a stringy exclusion principle,''
  JHEP {\bf 9812}, 005 (1998)
  [arXiv:hep-th/9804085].
}

\lref\Hohn{ G.~Hoehn, ``Selbstduale Vertexoperatorsuperalgebren und
das Babymonster,'' arXiv:0706.0236. }

\lref\Tuiteb{ G.~Mason and M.~P.~Tuite, ``On Genus Two Riemann
Surfaces Formed from Sewn Tori,'' arXiv:math.qa/0603088. }

\lref\ManschotZB{
  J.~Manschot,
  ``$AdS_3$ Partition Functions Reconstructed,''
  arXiv:0707.1159 [hep-th].
}

\lref\GaiottoXH{
  D.~Gaiotto and X.~Yin,
  ``Genus Two Partition Functions of Extremal Conformal Field Theories,''
  arXiv:0707.3437 [hep-th].
}

\lref\KrasnovZQ{
  K.~Krasnov,
  ``Holography and Riemann surfaces,''
  Adv.\ Theor.\ Math.\ Phys.\  {\bf 4}, 929 (2000)
  [arXiv:hep-th/0005106].
}

\lref\TakZog{
  P.~G.~Zograf and L.~A.~Takhtajan,
  ``On the Uniformization of Riemann Surfaces and on the Weil-Petersson Metric
  on the Teichm\"uller and Schottky Spaces,''
  Math.USSR-Sb.{\bf 60} (1988), no. 2, 297-313.
}

\lref\McIntyreXS{
  A.~McIntyre and L.~A.~Takhtajan,
  ``Holomorphic factorization of determinants of laplacians on Riemann surfaces
  and a higher genus generalization of Kronecker's first limit formula,''
  arXiv:math/0410294.
}

\lref\LeblancYY{
  Y.~Leblanc,
  ``THE GENUS 2 FREE ENERGY OF THE CLOSED BOSONIC STRING,''
  Phys.\ Rev.\  D {\bf 39}, 3731 (1989).
}

\lref\CoussaertZP{
  O.~Coussaert, M.~Henneaux and P.~van Driel,
  ``The Asymptotic dynamics of three-dimensional Einstein gravity with a
  negative cosmological constant,''
  Class.\ Quant.\ Grav.\  {\bf 12}, 2961 (1995)
  [arXiv:gr-qc/9506019].
}

\lref\TakhtajanMD{
  L.~A.~Takhtajan and L.~P.~Teo,
  ``Quantum Liouville theory in the background field formalism. I: Compact
  Riemann surfaces,''
  Commun.\ Math.\ Phys.\  {\bf 268}, 135 (2006)
  [arXiv:hep-th/0508188].
}

\lref\GaberdielVE{
  M.~R.~Gaberdiel,
  ``Constraints on extremal self-dual CFTs,''
  arXiv:0707.4073 [hep-th].
}

\lref\FriedanUA{
  D.~Friedan and S.~H.~Shenker,
  ``The Analytic Geometry of Two-Dimensional Conformal Field Theory,''
  Nucl.\ Phys.\  B {\bf 281}, 509 (1987).
}

\lref\DijkgraafFQ{
  R.~Dijkgraaf, J.~M.~Maldacena, G.~W.~Moore and E.~P.~Verlinde,
  ``A black hole farey tail,''
  arXiv:hep-th/0005003.
}

\lref\maloneywitten{
  A.~Maloney and E.~Witten,
  ``Quantum Gravity Partition Functions in Three Dimensions,''
  arXiv:0712.0155 [hep-th].
}

\Title{\vbox{\baselineskip12pt\hbox{} }} {\vbox{\centerline{Partition Functions
of Three-Dimensional Pure Gravity}}} \centerline{ Xi Yin }
\smallskip
\centerline{Jefferson Physical Laboratory, Harvard University,
Cambridge, MA 02138} \vskip .6in \centerline{\bf Abstract} { The
three-dimensional pure quantum gravity with a negative cosmological
constant has been conjectured to be dual to an extremal conformal
field theory (ECFT), of central charge $c=24k$ for some positive
integer $k$. We compute the partition function of the dual ECFT
by summing over gravitational instanton
contributions. In particular, we conjecture an exact expression for the contribution
from handlebodies to the partition function for all genera and all values of $k$, and provide nontrivial
evidences for the conjecture at genus two. } \vskip .3in

\Date{October 2007}
\listtoc \writetoc \noblackbox

\newsec{Introduction}

It is an interesting problem to understand whether three-dimensional
pure gravity exists as a quantum theory, and to solve it, if it
exists. Witten \WittenKT\ argued that three-dimensional pure gravity
with a negative cosmological constant in $AdS_3$ should be dual to
an ``extremal" conformal field theory (ECFT) on the boundary. The
ECFT factorizes into a holomorphic CFT and an anti-holomorphic CFT,
and we will mostly consider the holomorphic sector of the CFT. The
ECFT has central charge $c=24k$, where $k$ is a positive integer. So
far the only case where the ECFT is known to exist is $k=1$ \FLM.
Evidences for the existence of $k=2$ ECFT were found in \WittenKT\
by showing that its partition function on any hyperelliptic Riemann
surface can be consistently constructed. It is further shown in
\GaiottoXH\ that the genus two partition function of the $k=3$ ECFT
can be consistently constructed as well.\foot{On the other hand, arguments against the
existence of ECFT for large $k$ were presented in \GaberdielVE,
based on a conjectural differential equation that constrains the
genus one partition function. }

Naturally, one may ask whether the partition function of an ECFT of
general $k$, say on a Riemann surface of genus two, can be
constructed consistently and whether the answer is unique. One
may also wonder to what extent these partition functions can be
reproduced from a gravity computation, assuming that the ECFTs are
indeed dual to pure quantum gravity in $AdS_3$. In this paper we
will attempt to address both questions, and find evidence suggesting
that the partition function
of the ECFT (when it exists) can be produced exactly from the gravity
path integral, as a sum over contributions from gravitational instantons.
Moreover, we will
conjecture an exact expression for contribution from handlebodies
in pure three-dimensional gravity of all values of $k$, which is
explicitly computable (at least when the Riemann surface is hyperelliptic).
Our results suggest that the contributions from the handlebodies could dominate the gravity
path integral, and are in particular responsible for the ``polar" part of
the full partition function.


This paper is organized as follows. Section 2 gives an overview of
our main results and conjectures. The computation of the classical
gravity partition function is described in section 3. In section 4
we will discuss the $1/k$ corrections in the gravity computation,
and relate them to the ECFT partition function. Section 5
generalizes our proposals to higher genera. In section 6, we
describe some possible non-handlebody contributions. Details of the
computations, as well as our conventions for Siegel modular forms,
Schottky parameterizations, and the sewing/cutting of genus two
Riemann surfaces, are described in the appendices.

\newsec{The genus two partition function: gravity vs CFT}

We will now describe some general properties of the genus two
partition functions of the $c=24k$ ECFT, as well as its expected
relation to the gravity partition function. The genus two partition
function of a CFT with nonzero central charge is subject to the
conformal anomaly. When the CFT is holomorphic, we can require the
partition function to vary holomorphically with the moduli of the
Riemann surface. In the case of genus two, further requiring
$Sp(4,{\bf Z})$ modular invariance fixes the partition function up
to an overall constant. More precisely, the genus two partition
function of the $c=24k$ ECFT can be regarded as a
Siegel modular form of weight $2k$,\foot{More precisely, we allow these modular forms
to have poles along the divisor in the Siegel upper half space corresponding to
the separating degeneration of the Riemann surface; we will loosely call these Siegel
modular forms as well.} denoted by
$Z_{k,g=2}^{mod}(\Omega)$, or $Z_k^{mod}(\Omega)$ for short. This is
the partition function considered in \WittenKT\ and \GaiottoXH. One
should be cautious that \Tuite, for instance, uses a different genus
two partition function, related to $Z_{k,g=2}^{mod}$ by a
``holomorphic correction" factor $G(\Omega)^{-k}$, which is not a
Siegel modular form. This difference would be important if one
studies the factorization of the partition function at the
separating degeneration using the $\epsilon$-parameter of \Tuite.
The explicit expression for $Z_{k,g=2}^{mod}(\Omega)$ with $k=1$ was
obtained in \Tuite, whereas the $k=2,3$ results were computed in
\GaiottoXH. It is of the form \eqn\offs{ Z_{k,g=2}^{mod}(\Omega) =
{T_k(\Omega)\over \chi_{10}^k} } where $\chi_{10}$ is the weight 10
Igusa cusp form, and $T_k(\Omega)$ is an entire Siegel modular form
of weight $12k$, given as a polynomial in the generating forms
$\psi_4,\psi_6,\chi_{10},\chi_{12}$.

The gravity partition function, on the other hand, is computed by
summing over saddle point contributions to the path integral. Each
saddle point corresponds to a classical solution to the Euclidean
equation of motion, i.e. a hyperbolic three-manifold $M$ whose conformal
boundary is the given Riemann surface $\Sigma$. In this paper, we will be
mostly considering a particularly
simple class of such hyperbolic three-manifolds, namely the ones which are
handlebodies.\foot{A (3-dimensional) handlebody
is homeomorphic to the domain enclosed by a closed surface embedded in ${\bf R}^3$.}
In general, when $\Sigma$ has genus $g>1$, there are a lot of hyperbolic three-manifolds whose conformal boundary is $\Sigma$,
that are not handlebodies;\foot{I'm grateful to E. Witten for pointing this out.} we
do not understand their contributions and will comment on them at the end.

The classical instanton action is given by a suitably regularized
Einstein-Hilbert action evaluated on $M$. In addition,
there are quantum corrections, suppressed by powers of the coupling
constant $1/k$. Higher than 2-loop corrections will vanish if the
boundary Riemann surface has genus one, but are in general
non-vanishing for higher genus Riemann surfaces.

\vskip 0.5cm
\centerline{\vbox{\centerline{ \hbox{\vbox{\offinterlineskip
\halign{&#&\strut\hskip0.2cm \hfill #\hfill\hskip0.2cm\cr
\epsfysize=1.3in \epsfbox{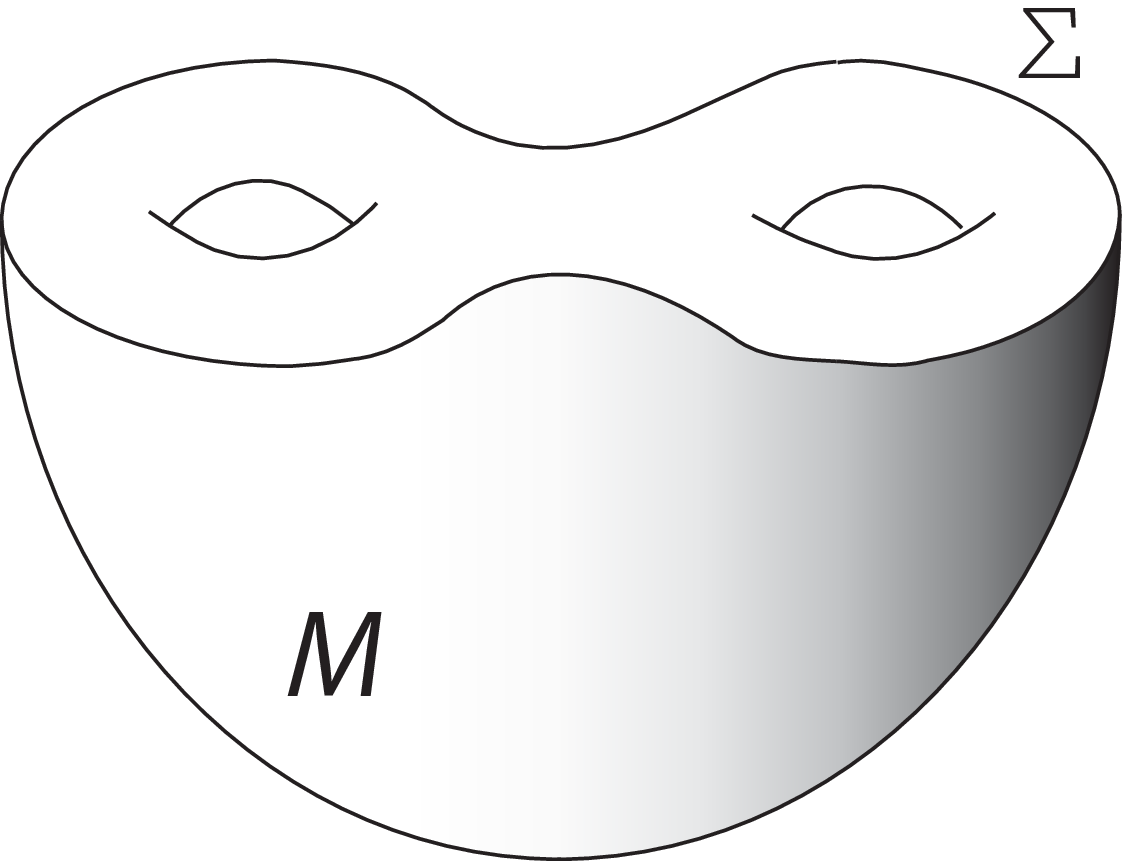}  \cr }}}}
{{\bf Figure 1.} The Riemann surface $\Sigma$ as the conformal boundary
of a hyperbolic three-manifold $M$.}
}} \vskip 0.5cm

The contribution from the path integral around a saddle point,
corresponding to a handlebody $M$,
takes the form \eqn\aafulson{ Z_{saddle}(k,\Omega)=\exp\left[ k
S_0(\Omega) + S_1(\Omega) + {1\over k} S_2(\Omega) + {1\over k^2}
S_3(\Omega)+\cdots \right] } where $k^{1-l}S_l(\Omega)$ is the
$l$-loop free energy of the boundary graviton excitations. $\Omega$ is the
period matrix of the Riemann surface; in the genus two case, we write
\eqn\peirdm{ \Omega = \pmatrix{
\rho & \nu \cr \nu & \sigma }. }
 The tree
level (classical) contribution is given by \eqn\szero{ e^{S_0} =
{{\cal F}(\Omega)^{12}\over \chi_{10}(\Omega)} } where ${\cal
F}(\Omega)$ is a function of the moduli of the genus two Riemann
surface, together with the choice of cycles that are contractible in
the hyperbolic three-manifold. Its precise expression will be
defined later. We will also conjecture an explicit formula for
$S_1$.

The contribution from all handlebodies to the full gravity partition
function is given by summing over
$Sp(4,{\bf Z})$ images of \aafulson, with weight $2k$. We expect
$Z_{saddle}$ to be invariant under $\Gamma_\infty\subset Sp(4,{\bf
Z})$; this is in particular true for ${\cal F}(\Omega)$. So the total
handlebody contribution is \eqn\aaexactpf{ Z_{h.b.}(k,\Omega)=\sum_{\gamma\in
\Gamma_\infty\backslash Sp(4,{\bf Z})} \det(C\Omega+D)^{-2k}
Z_{saddle}(k,\gamma\cdot\Omega) }
The sum in \aaexactpf\ converges
if $Z_{saddle}(k,\Omega)$ satisfies suitable regularity conditions, and gives an $Sp(4,{\bf Z})$
Siegel modular form. Furthermore, suitable ``polar"
terms of $Z_{h.b.}(k,\Omega)$ come entirely from the first term in the sum in \aaexactpf, i.e. $Z_{saddle}(k,\Omega)$
itself. We expect \aaexactpf, together with possible non-handlebody contributions,
to give exactly the modular ECFT partition function
$Z_{k,g=2}^{mod}(\Omega)$.

It is of interest to compute the $S_l$'s ($l\geq 1$) directly from
loops of boundary graviton excitations. We will however adopt an
alternative approach. The idea is that, the boundary graviton
excitations are nothing but Virasoro descendants of the vacuum,
propagating along the handles of the boundary Riemann surface which
are filled in by the hyperbolic three-manifold.
 Let us
consider a ``fake" CFT partition function \eqn\fakecfp{\eqalign{
Z_{fake}(k,\Omega) &= G(\Omega)^{k}\sum_{{\cal A}_i\in Vir(k)}
\epsilon^{\Delta_i-2k} {\rm Tr}_{Vir(k)}({\cal A}_i e^{2\pi
i\tau_1(L_0-k)}) {\rm Tr}_{Vir(k)}({\cal A}_i e^{2\pi
i\tau_2(L_0-k)})  } } where $\tau_1$ and $\tau_2$ are the moduli of
two tori, glued together to form the genus two Riemann surface, with
$\epsilon$ being the pinching parameter. The precise definition of
$\tau_1,\tau_2$ and $\epsilon$ are given in \Tuite, which are
functions of the period matrix $\Omega$. In the separating
degeneration limit $\epsilon\to 0$, $\tau_1,\tau_2,\epsilon$ are
approximately $\rho,\sigma,2\pi i\nu$. $G(\Omega)$ is the universal
``holomorphic correction" factor of \Tuite, needed in relating the
torus one-point functions to the modular genus two partition
function. The ${\cal A}_i$'s in \fakecfp\ run through a set of basis
operators in a $c=24k$ Virasoro algebra, orthonormal with respect to
the Zamolodchikov metric. The traces are over all Virasoro
descendants of 1. If ${\cal A}_i$'s were to run over all operators
in a unitary CFT, and the traces in \fakecfp\ over all operators,
\fakecfp\ would have given the modular genus two partition function.

Our main conjecture is that, $Z_{fake}$ is in fact the same as
the contribution $Z_{saddle}$ from the gravitational instanton corresponding to a
handlebody. $Z_{fake}$ is expected to approximate the ECFT partition function in
its expansion near $\tau_1,\tau_2\to i\infty$, $\epsilon\to 0$. In
fact, it captures all the terms of $Z_k^{mod}$ that are polar (and constant) in
$q=e^{2\pi i\rho},s=e^{2\pi i\sigma}$ (or equivalently, in
$e^{2\pi i\tau_1}, e^{2\pi i\tau_2}$), since these correspond to the
traces in \fakecfp\ over operators with $L_0\leq k$, and there are
no nontrivial primaries of such weights in the ECFT; ${\rm
Tr}_{Vir(k)} {\cal A} q^{L_0-k}$ would be non-vanishing only if
${\cal A}$ is a Virasoro descendant of 1. In general, such polar
terms do not entirely determine $Z_k^{mod}$, since there is an
ambiguity of adding to $Z_k^{mod}$ weight $2k$ cusp forms (i.e.
vanishing as $q\to 0$ or $s\to 0$), for $k\geq 5$. For example,
the correlation function of primaries $\langle {\cal
O}_m {\cal O}_i {\cal O}_m \rangle$ contributes to $Z_k^{mod}(\Omega)$,
in the factorization limit, a term of order $e^{2\pi i
(\Delta_m-k)(\tau_1+\tau_2)}\epsilon^{\Delta_i-2k} \sim
q^{\Delta_m-k} s^{\Delta_m-k}\nu^{\Delta_i-2k}$. In general such terms are not
fixed by the polar terms, and may not be reproduced by summing
over $\Gamma_\infty \backslash Sp(4,{\bf Z})$ images of $Z_{fake}$.


To summarize, our conjectures (in the genus two case) are:

{\bf (1)} The saddle point contribution $Z_{saddle}$ in the gravity
path integral, corresponding to a handlebody, takes the form
\aafulson, with $S_0$ given by \szero\ (there is also an explicit
conjectural formula for $S_1$, to be described later.)

{\bf (2)} $Z_{saddle}(k,\Omega)=Z_{fake}(k,\Omega)$ for all $k$, the
latter being defined by \fakecfp\ and explicitly computable.

{\bf (3)} The dual $c=24k$ ECFT modular genus two partition function
$Z_k^{mod}(\Omega)$ is given by summing over the gravitational
instanton contributions, handlebodies as well as non-handlebodies in
general. The handlebody contribution $Z_{h.b.}$ \aaexactpf\ gives
all the polar terms of $Z_k^{mod}$.

These conjectures also admit straightforward generalizations to
higher genera. In the remaining sections, we will further explain
them and present partial evidences. In section 3.4 and 4.1, we will
match the terms in $Z_{saddle}(k,\Omega)$ that are of non-positive
powers in $q=e^{2\pi i\rho}$, $s=e^{2\pi i\sigma}$ and to order
${\cal O}(\nu^{4-2k})$, with those in $Z_k^{mod}(\Omega)$, in the
cases $k=1,2,3$. Our most nontrivial checks are presented in section
4.3, in which our conjectured formulae for $S_0$ and $S_1$ are shown
to agree with the leading terms in the $1/k$ expansion of $\ln
Z_{fake}(k,\Omega)$ up to order $\nu^4$.

\newsec{Classical partition function of pure 3d gravity}

\subsec{Genus one}

We shall consider the geometry of a Euclidean hyperbolic 3-manifold
with an asymptotic boundary that is conformally equivalent to a
torus with complex modulus $\tau$. The regularized Einstein-Hilbert
action of this solution is known to be \MaldacenaBW\ \eqn\aa{ S =
{4\pi k}\tau_2 = -2\pi ik(\tau-\bar\tau) } where $c=24k=3l/2G$, $l$
being the curvature radius and $G$ being Newton's constant. The
classical limit of the path integral should sum over different ways
of filling in the boundary torus, parameterized by $\Gamma_\infty
\backslash SL(2,{\bf Z})$. So the ``naive" genus one partition
function is \eqn\gnto{ \sum_{\gamma\in \Gamma_\infty \backslash
SL(2,{\bf Z})} \exp\left[-2\pi ik({a\tau+b\over
c\tau+d}-{a\bar\tau+b\over c\bar\tau+d})
\right],~~~~~\gamma=\pmatrix{a & b\cr c & d}. } This result is
however in conflict with the assumption of holomorphic
factorization. In fact, there appears to be no reason why \gnto\
would have an expansion of the form $\sum a_{n,m}q^n \bar q^m$ with
integer coefficients $a_{n,m}$, thus in conflict with its
interpretation as the partition function of discrete states in
$AdS_3$.\foot{This is pointed out to me by S. Minwalla. It is also
observed by \maloneywitten.} We will assume that the left and right
moving sectors, corresponding to Chern-Simons gauge fields $A_L$ and
$A_R$ of the $SO(2,1)\times SO(2,1)$, contribute to the partition
function independently. We must then include more general classical
Euclidean solutions that involve complex metrics, and the partition
function takes the form $|Z_{g=1}(\tau)|^2$, where \eqn\dtnit{
Z_{g=1}(\tau) = \sum_{\gamma\in \Gamma_\infty \backslash SL(2,{\bf
Z})} \exp\left[-2\pi ik({a\tau+b\over c\tau+d}) \right] } This sum
is apparently divergent. It nevertheless can be regularized, as
explained in \ManschotZB, and gives rise to the weakly holomorphic
$SL(2,{\bf Z})$ modular form whose only polar term in its
$q$-expansion is $q^{-k}$. Up to corrections due to boundary
excitations of gravitons (which are one-loop in $1/k$),
$Z_{g=1}(\tau)$ coincides with the genus one partition function of
the extremal CFT with $c=24k$.

\subsec{Genus two and higher}

For genus $g>1$, we must compute the regularized Einstein-Hilbert
action of the hyperbolic three-manifold $M$ whose conformal boundary
is the given genus $g$ Riemann surface $\Sigma_g$. In general, $M$
can be constructed as the quotient of hyperbolic 3-space ${\bf H}_3$
by a Kleinian group $\Gamma\subset SL(2,{\bf C})$. When $M$ is a
handlebody, the construction is particularly simple; $\Gamma$ will
be a Schottky group, i.e. a freely (finitely) generated subgroup of
$SL(2,{\bf C})$ that is purely loxodromic. The conformal boundary of
${\bf H}_3$ is a ${\bf CP}^1$. $\Gamma$ is freely generated by
$\gamma_1,\cdots,\gamma_g\in SL(2,{\bf C})$, which act on ${\bf
CP}^1$ as mobius transformations. The quotient of ${\bf P}^1$ (minus
a zero measure limiting set) by $\Gamma$ gives the Riemann surface
$\Sigma_g$. There are $3g$ complex parameters for
$\gamma_1,\cdots,\gamma_g$, but they are equivalent to their
conjugations by the overall $SL(2,{\bf C})$. So there are $3g-3$
independent complex parameters, agreeing with the number of complex
moduli of the Riemann surface. The $3g-3$ complex variables
parameterizing the generators of $\Gamma$ are coordinates on the
Schottky space, which is a covering space of the moduli space of
$\Sigma_g$.

\vskip 0.5cm
\centerline{\vbox{\centerline{ \hbox{\vbox{\offinterlineskip
\halign{&#&\strut\hskip0.2cm \hfill #\hfill\hskip0.2cm\cr
\epsfysize=1.6in \epsfbox{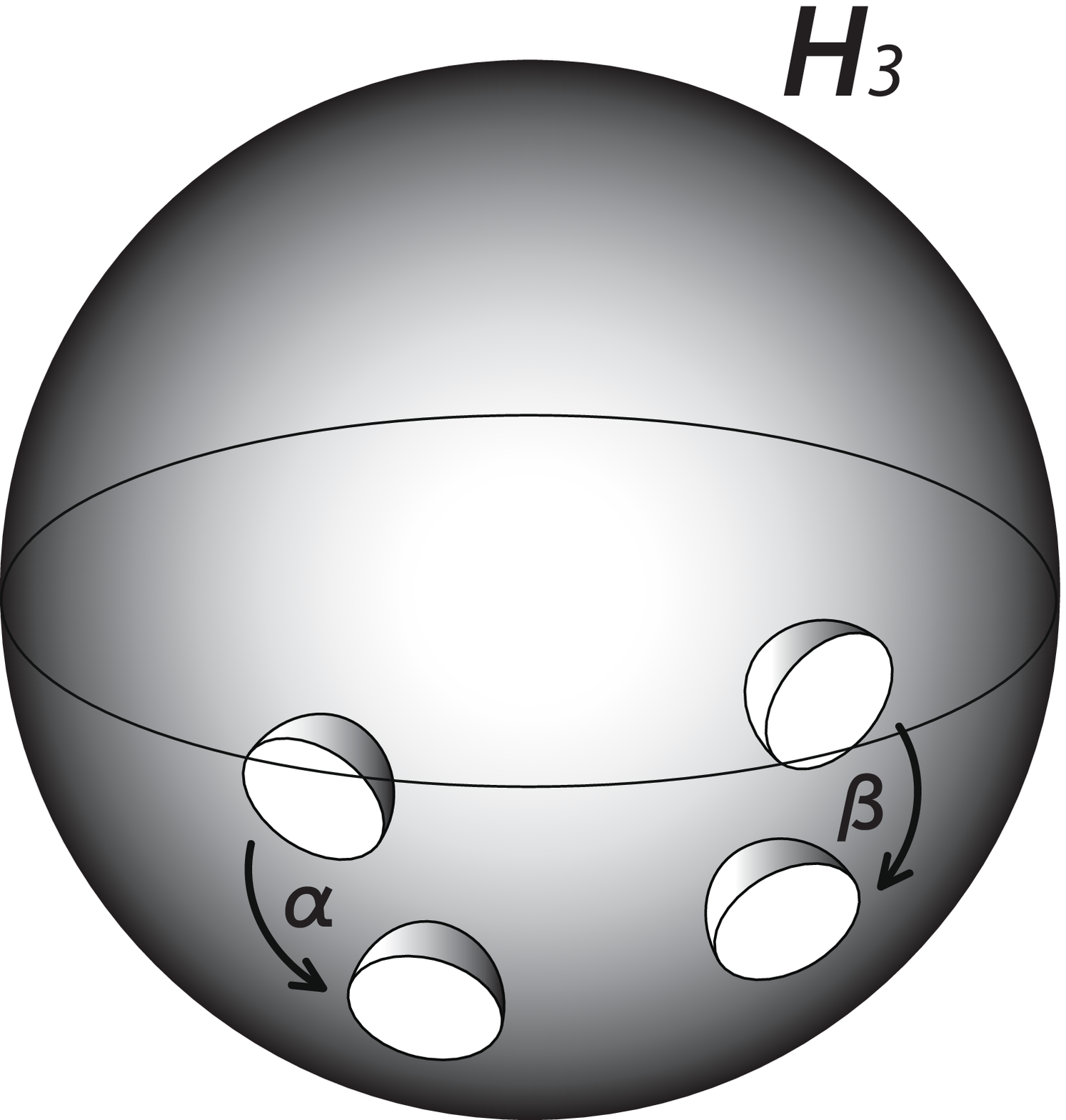}  \cr }}}}
{{\bf Figure 2.} A genus two handlebody $M$ represented as the quotient
of hyperbolic 3-space ${\bf H}_3$ by the Schottky group with two generators $\alpha,\beta$. The shaded region
(outside the four hemispheres) is a fundamental domain for $M$.}
}} \vskip 0.5cm

It is shown in \KrasnovZQ\ that the regularized Einstein-Hilbert
action evaluated on the hyperbolic three-manifold is given by a
suitably defined classical Liouville action $S_L[\phi]$ evaluated at
its critical point (i.e. $ds^2=e^\phi dzd\bar z$ being the
uniformizing metric), whose value we denote by $S_L$. An important
point is that, the Liouville action cannot be defined by naively
integrating the Liouville Lagrangian density ${\cal L}=(\partial
\phi)^2 + e^\phi$ over the entire Riemann surface, as ${\cal L}$
does not transform covariantly under coordinate transformations
between differences patches of the Riemann surface. The way to fix
this is to define the Liouville action by integrating the Lagrangian
density over a fundamental domain in ${\bf C}$ that parameterizes
the Riemann surface $\Sigma_g$, with suitable boundary terms
included \TakZog.

There are two standard parameterizations of $\Sigma_g$, the Schottky
parameterization and Fuchsian parameterization. The former, as
described earlier, models $\Sigma_g$ as a quotient of the complex
plane ${\bf C}$ (or ${\bf P}^1$) by the Schottky group
$\Gamma\subset SL(2,{\bf C})$. The latter models $\Sigma_g$ as the
quotient of the hyperbolic plane by a discrete subgroup of
$SL(2,{\bf R})$. It was shown in \TakZog\ that the Liouville action
defined using the Schottky parameterization evaluates at its
critical point to a K\"ahler potential of the Weil-Petersson metric
on the Schottky space, whereas the action defined in Fuchsian
parameterization evaluates to a trivial constant proportional to
$2g-2$. We are looking for a suitably defined $S_L$ that is the sum
of a holomorphic function and an anti-holomorphic function in the
moduli, and neither of these could serve as the classical saddle
point contribution in three-dimensional pure gravity.

A hint comes from the following formula for the holomorphic
factorization of the determinant of the scalar Laplacian on
$\Sigma_g$, due to Zograf and was described in \McIntyreXS:
\eqn\zogr{ {\det \Delta\over \det {\rm Im}\Omega} = c_g e^{-{1\over
12}S_L(\Omega)} |{\cal F}(\Omega)|^2 } On the RHS of \zogr, $c_g$ is
a constant that depends on $g$ only. In the context of \McIntyreXS,
the determinant of the Laplacian is defined using zeta function
regularization in the uniformizing hyperbolic metric on $\Sigma_g$
(with constant curvature $-1$), and $S_L(\Omega)$ is the Liouville
action defined in Schottky parameterization. ${\cal F}(\Omega)$ is a
holomorphic function on the Schottky space, given by the following
infinite product formula \eqn\dotn{ {\cal F}(\Omega) =
\prod_{\gamma~ prim.} \prod_{m=1}^\infty (1-q_\gamma^m) } where the
first product runs over primitive conjugacy classes of the Schottky
group $\Gamma=\langle\gamma_1,\cdots,\gamma_g\rangle$. Primitive
here means that $\gamma$ is not a positive power of any other
element in $\Gamma$. $q_\gamma$ is defined as follows: every element
$\gamma\in\Gamma$ is (uniquely) conjugate under $SL(2,{\bf C})$ to
$z\mapsto \lambda z$, with $|\lambda|<1$, and $q_\gamma\equiv
\lambda$. More explicitly, \eqn\fixabc{ \eqalign{
{\gamma(z)-\eta\over \gamma(z)-\xi} = q_\gamma {z-\eta\over z-\xi}.
} } where $\xi,\eta$ are fixed points of $\gamma$. Since the
Schottky group elements $\gamma$ in fact only depend on the period
matrix $\Omega$ modulo integral shifts, $q_\gamma$ are functions of
$\{e^{2\pi i \Omega_{mn}}\}$ (but they are not simply products of
$e^{2\pi i \Omega_{mn}}$).

To be explicit, let us restrict to the genus two case. In general
one expects an ambiguous factor in the determinant of the Laplacian
due to the conformal anomaly -- there is no canonical choice of the
scale factor of the metric on the conformal boundary $\Sigma_g$; but
this is the same kind of ambiguity that appears in $S_L$. To
determine the Liouville action $S_L$ corresponding to the classical
gravitational instanton action, consider the modular partition
function of a single free boson, \eqn\tot{ (\det \Delta)^{-\half} =
(\det{\rm Im}\Omega)^{-\half} |\chi_{10}(\Omega)|^{-{1\over 12}}. }
Plugging \tot\ into \zogr, we obtain a holomorphically factorized
``Liouville action" $S_L=S_{hol}+\overline{S_{hol}}$, with
\eqn\jotl{ e^{-{1\over 12}S_{hol}} = \chi_{10}(\Omega)^{1\over 12}
{\cal F}(\Omega)^{-1} } The holomorphic part of the Euclidean action
for pure three-dimensional gravity of $c=24k$ should then be
identified with \eqn\tkst{ e^{{k}S_{hol}} = \left[{ {\cal
F}(\Omega)^{12} \over \chi_{10}(\Omega)}\right]^k } In the end, we
want to sum over all inequivalent $Sp(4,{\bf Z})$ images of \tkst,
which should produce the contribution to the partition function from
all handlebodies, up to loop corrections suppressed by $1/k$. In
general, the handlebody filling in a Riemann surface $\Sigma_g$ is
invariant under $\Gamma_\infty\subset Sp(2g,{\bf Z})$, the subgroup
of $Sp(2g,{\bf Z})$ that fixes the cusp $\Omega =i\infty\cdot {\bf
1}$. Equivalently, $\Gamma_\infty$ consisting of elements of the
form \eqn\tstatg{ \gamma=\pmatrix{A & B\cr 0 & D}\in Sp(2g,{\bf
Z}),~~~~AD^T=1,~~~AB^T=BA^T. } Thus we expect ${\cal F}(\Omega)$ to
be invariant under $\Gamma_\infty$. This is indeed the case, as
shown in appendix B.

\subsec{The expansion of ${\cal F}(\Omega)$}

The expression \dotn\ for ${\cal F}$ is rather complicated. In
practice, we would like to expand ${\cal F}$ near the separating
degeneration limit $\nu\to 0$, and compare with the corresponding
expansion of a Siegel modular form. Firstly, to compute $q_\gamma$
one must express the Schottky parameters in terms of the periods
$\rho,\sigma,\nu$. An explicit formula for this is given in
\LeblancYY\ and is recalled in Appendix B. Next, one can order the
conjagacy classes $\gamma$ according to the order of $q_\gamma$ in
$\nu$. The primitive conjugacy classes are
\eqn\prims{\alpha^{\pm1},~~~\beta^{\pm1},~~~\alpha^n\beta^m~(n,m\not=0),~~~
\alpha^n\beta^m\alpha^{n'}\beta^{m'}~((n,m)<(n',m')),~~~\cdots\cdots}
where for $\alpha^n\beta^m\alpha^{n'}\beta^{m'}$ we must choose
$(n,m)$ to be distinct from $(n',m')$; and interchanging $(n,m)$
with $(n',m')$ gives the same conjugacy classes. In the $\nu\to 0$
limit, we shall find \eqn\sinf{ q_{\alpha^\#},q_{\beta^\#}\sim {\cal
O}(1),~~~~q_{\alpha^\#\beta^\#}\sim \nu^2,~~~~
q_{\alpha^\#\beta^\#\alpha^\#\beta^\#}\sim \nu^4,~~~~{\rm etc.} }
where $\#$ stands for arbitrary nonzero integers. Up to order
$\nu^4$, it suffices to take into account only the conjugacy classes
in \prims\ in the product formula for ${\cal F}(\Omega)$. The
details of the computation can be found in appendix C. The result is
\eqn\foform{ \eqalign{ & {\cal F}(\Omega)= \prod_{m=1}^\infty
(1-q^m)^2 (1-s^m)^2 \left\{ 1+ (2\pi i\nu)^2 4 \hat E_2^\rho \hat
E_2^\sigma +{(2\pi i\nu)^4\over 3} \left[-2 (\hat E_2^\rho)^2 \hat
E_2^\sigma - 2 \hat E_2^\rho (\hat E_2^\sigma)^2 \right. \right.\cr
&\left.\left. ~~~~~~~~~~~~~~~~~~~~+48 (\hat E_2^\rho)^2(\hat
E_2^\sigma)^2-10 (\hat E_2^\rho)^2 \hat E_4^\sigma - 10 (\hat
E_2^\sigma)^2 \hat E_4^\rho-5 \hat E_4^\rho \hat E_4^\sigma \right]
+ {\cal O}(\nu^6) \right\} } } where $\hat E_n^\rho$ is the $n$-th
Eisenstein series $E_n(\rho)$ with the constant term subtracted, and
normalized so that its $q$-expansion starts with $q+\cdots$. In
other words, $\hat E_n^\rho = \sum_{m=1}^\infty {m^{n-1}q^m\over
1-q^m}$, and similarly for $\hat E_n^\sigma$.

\subsec{Series expansion of the ECFT genus two partition functions}

In order to compare the classical gravity partition function with the known ECFT genus two
partition functions, we shall consider the Laurent expansion of the latter in
$q,s$ and $\nu$. $Z_{k,g=2}^{mod}$ with $k=1,2,3$ are given explicitly in terms of the
generating Siegel modular forms $\psi_4,\psi_6,\chi_{10},\chi_{12}$
in \refs{\Tuite,\GaiottoXH}. The generating forms can be expressed
in terms of products and/or sums of the 10 weight
$\half$ characteristic theta series, as in Appendix A. We find the
following expansions of the $T_k(\Omega)$'s (defined in
\offs), to order $(q^k,s^k,\nu^4)$:
\eqn\texps{ \eqalign{ & T_1(\Omega) = (1-24q)(1-24s)+48 qs(2\pi
i\nu)^2-20 qs(2\pi i\nu)^4 + {\cal O}(q^2,s^2,\nu^6) \cr &
T_2(\Omega) = (1-48 q+1081 q^2)(1-48 s+1081
s^2)+{qs(-24+1081q)(-24+1081s)\over 6}(2\pi i\nu^2)\cr &~~~~+
{qs\left[-2880+105384(q+s)-3478739 qs\right]\over 72}(2\pi i\nu)^4 +
{\cal O}(q^3,s^3,\nu^6) \cr & T_3(\Omega) = (1 - 72 q + 2485 q^2 -
54599 q^3)(1 - 72 s + 2485 s^2 - 54599 s^3)\cr &~~~~+ {q s(-72 +
4970 q - 163797 q^2)(-72 + 4970 s - 163797 s^2)\over 36} (2\pi
i\nu)^2 \cr &~~~~+qs\left[-60 + {10907\over 3}(q+s) - {204729\over2}
(q^2+s^2)\right.\cr &~~~~~~\left.  - {22571123\over 108} q s +
{130929005\over 24} q s(q+s) - {2004085311\over 16} q^2
s^2\right](2\pi i\nu)^4 + {\cal O}(q^4,s^4,\nu^6) } } The higher order terms in $q$
and $s$ are
non-polar and are not expected to agree with $Z_{saddle}$. Indeed, for $k=1,2,3$, we
find that $T_{k}$ agrees with ${\cal F}(\Omega)^{12k}$ up to
terms of order ${\cal O}(q^2,s^2,\nu^6)$. In the case $k=1$, this is all the agreement one could
hope for. For $k=2,3$, at
order $q^2,s^2$ and higher, one must include the $1/k$ loop corrections $S_1, S_2,\cdots$,
as will be discussed in the next section.

\newsec{Going beyond the classical level}

\subsec{The loop expansion in $1/k$}

We expect that the full (perturbative) contribution around a saddle
point, $Z_{saddle}$, corresponding to a particular filling
hyperbolic three-manifold $M$, takes the form \aafulson.  When $M$
is a handlebody, $Z_{saddle}$ should be invariant under
$\Gamma_\infty$, and the summation over $Sp(4,{\bf
Z})/\Gamma_\infty$ images of \aafulson\ then account for all
handlebody instanton contributions. The bulk gravity Lagrangian
reduces to that of $SL(2,{\bf R})$ WZW model on the boundary Riemann
surface \CoussaertZP, which factorizes into a chiral part and an
anti-chiral part. In principle, one should compute the $l$-loop free
energy of the $SL(2,{\bf R})$ WZW model on the genus two Riemann
surface, which is expected to give the term $k^{1-l}S_l$ in
\aafulson. We will describe an alternative approach to compute the
loop corrections in the next subsection.

Motivated by the product formula for the
one-loop free energy of Liouville theory on a general Riemann surface \TakhtajanMD,
we conjecture the following ``holomorphic version" of the product formula as the one-loop free energy of the
chiral $SL(2,{\bf R})$ WZW model \eqn\soneli{ e^{S_1} = \prod_{\gamma ~{
prim.}} \prod_{m=2}^\infty (1-q_\gamma^m)^{-{\half}} } Note that the
product in \soneli\ is similar to the one in \dotn, but with a
different range of $m$, and raised to the power $-\half$. The power $\half$
is due to our convention, counting $\gamma$ and $\gamma^{-1}$ as distinct
primitive classes. It is conceivable
that \soneli\ has the interpretation as the determinant of a $\bar\partial$-like operator
on the Riemann surface, and it would be nice to understand this. Another check of
\soneli\ is that, it is consistent with the genus one answer in the factorization limit,
and is manifestly $\Gamma_\infty$ invariant.

Up to
order $\nu^4$, \soneli\ can be calculated using the expansions of
$q_\alpha, q_\beta$ and $q_{\alpha^n\beta^m}$ computed earlier.
Explicitly, we have \eqn\expanssone{ \eqalign{ & e^{S_1} =
\prod_{n=2}^\infty (1-q^n)^{-1} (1-s^n)^{-1}\times \left[
1-2qs(q+s+3qs)(2+3q+3s+8qs)(2\pi i\nu)^2 \right.  \cr &\left.
+{qs\over 6}\left(-2(q+s)+45
(q^2+s^2)+72qs+745(q^2s+qs^2)+3720q^2s^2\right) (2\pi i\nu)^4 +{\cal
O}(q^4,s^4,\nu^6)\right] } }

As remarked at the end of section 3, $S_1, S_2,\cdots$ are needed in
$Z_{saddle}(k,\Omega)$ in order to compare with the polar terms in
$Z_k^{mod}(\Omega)$ for $k=2,3$. Let us compare $\chi_{10}^k
Z_{saddle}(k,\Omega)$ with $T_k(\Omega)$ \texps. At order $\nu^0$,
the agreement is a trivial consequence of the genus one result and
the factorization of $Z_k^{mod}(\Omega)$. At order $\nu^2$, only one
and two-loop corrections, i.e. $S_1, S_2$, in addition to $S_0$, are
involved in $Z_{saddle}$. Having the explicit (conjectured)
expressions for $S_0$ \foform\ and $S_1$ \expanssone, the matching
of $Z_{saddle}(k=2)$ with the polar terms in $Z_{k=2}^{mod}$ already
fixes $S_2$ at order $\nu^2$, up to $q^2,s^2$. Further requiring
$Z_{saddle}(k=3)$ to agree with $Z_{k=3}^{mod}$ then provides a
nontrivial check for the expression of $S_2$ up to order
$q^2s^2\nu^2$. This is indeed the case. Using the comparison with
the polar terms in the $k=3$ case, we can further determine $S_2$ up
to order $q^3s^3\nu^2$.

At order $\nu^4$, we expect $S_1,S_2,S_3$ to contribute to
$\chi_{10}^kZ_{saddle}$.\foot{In fact, we expect all $S_l$'s to contribute
at this order. However, it appears that the contributions from $S_{l\geq 4}$
are of higher order in $q,s$. } By comparing with up to order $q^2s^2\nu^4$
terms in $T_2(\Omega)$ and $T_3(\Omega)$ (which are polar), we can
determine the order $q^2s^2\nu^4$ terms in $S_2$ and $S_3$. The
result is summarized in the following expansion
\eqn\sitlee{\eqalign{ & S_2(\Omega)=q^2s^2\left({1\over 3}+ {1\over
2} q+{1\over 2}s+{3\over 4}qs+\cdots\right) (2\pi i\nu)^2 \cr
&~~~~~~~~~~+ q^2s^2\left( {13\over 36} + {1\over 8}(q+s)-{45\over
16}qs +\cdots \right)(2\pi i\nu)^4 + {\cal O}(\nu^6).\cr &
S_3(\Omega) = {\cal O}(q^4s^4\nu^4,\nu^6) }}
It would be interesting to reproduce \sitlee\ directly from perturbative computations in the chiral
$SL(2,{\bf R})$ WZW model.

\subsec{The ``fake" CFT on the Riemann surface}

The genus one expression for $Z_{saddle}$ \eqn\geonean{ q^{-k}
\prod_{n=2}^\infty (1-q^n)^{-1}, } which is valid to all-loop in
$1/k$, suggests that the only states propagating along handles of
the Riemann surface are Virasoro descendants of the ground state. We
are led to hypothesize that $Z_{saddle}$ can be computed by
pretending that we have a $c=24k$ (non-unitary) CFT on the boundary
Riemann surface, whose only operators are Virasoro descendants of
the identity, namely $T, \partial T,\cdots$. The genus two partition
function $Z_{fake}$ of this ``fake" CFT can be computed by gluing
the four-point functions of all Virasoro descendants on the sphere.
Equivalently, by the factorization of tree level four-point function
into three-point functions, $Z_{fake}$ is given by \fakecfp.

\vskip 0.5cm
\centerline{\vbox{\centerline{ \hbox{\vbox{\offinterlineskip
\halign{&#&\strut\hskip0.2cm \hfill #\hfill\hskip0.2cm\cr
\epsfysize=1.6in \epsfbox{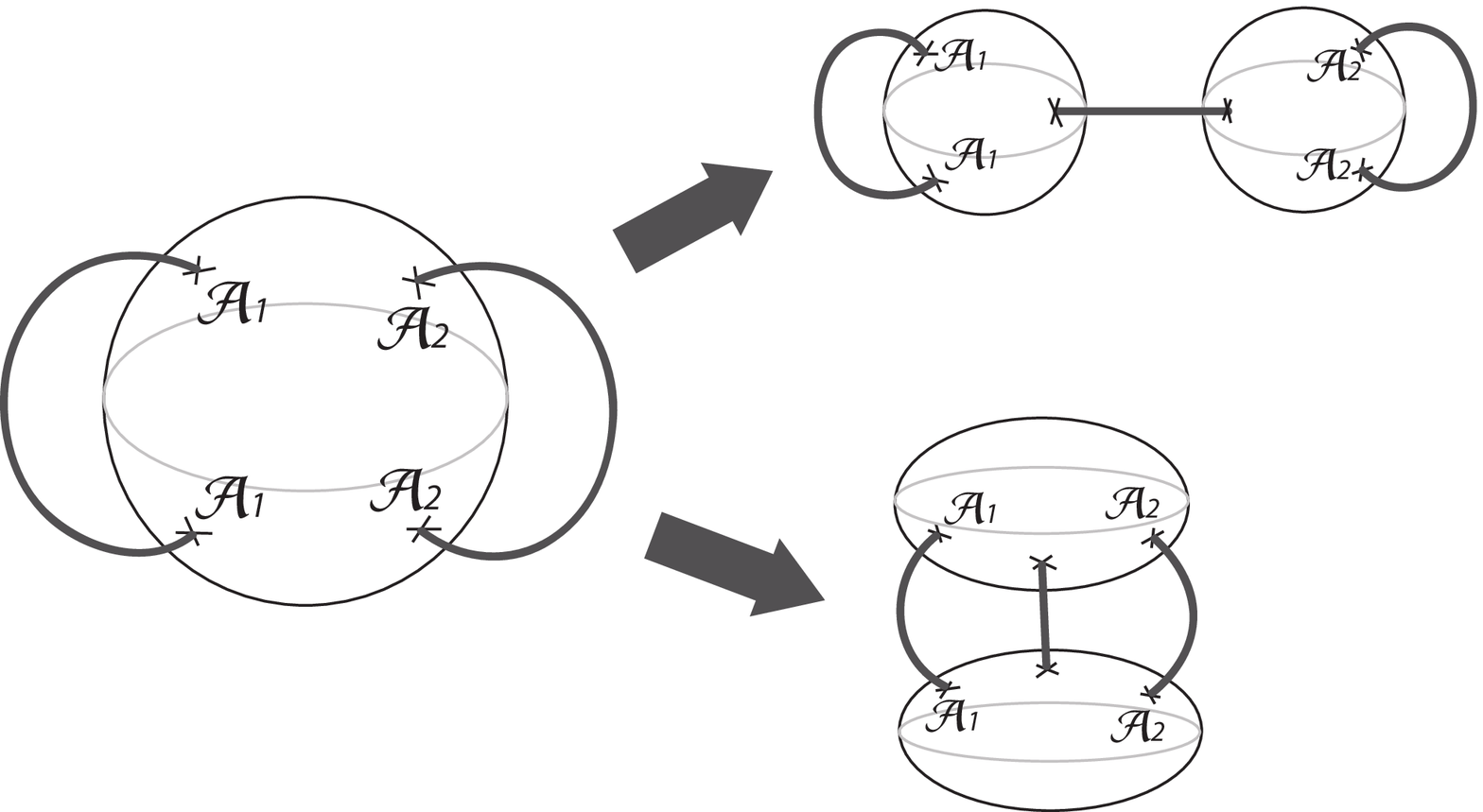}  \cr }}}}
{{\bf Figure 3.} The fake genus two partition function is obtained
by gluing sphere four-point functions of Virasoro descendants of the identity.
It can be factorized into three-point functions in two different ways.}
}} \vskip 0.5cm

An
alternative factorization of the four-point function of the Virasoro
descendants gives a representation of $Z_{fake}$ of the form
\eqn\altfake{ Z_{fake} = \tilde G(\Omega)^k \sum_{{\cal A}_i, {\cal
A}_j, {\cal A}_l\in Vir(k)} \langle {\cal A}_i {\cal A}_j {\cal A}_l
\rangle \tilde q_1^{\Delta_i-k} \tilde q_2^{\Delta_j-k} \tilde
q_3^{\Delta_l-k} \langle {\cal A}_i {\cal A}_j {\cal A}_l \rangle }
where $\tilde G(\Omega)$ is another possible universal ``holomorphic
correction" factor, and $\tilde q_i=e^{2\pi i\tilde\tau_i}$
($i=1,2,3$) are determined by the moduli of the genus two Riemann
surface. Another way to write \altfake\ is to express the fake
partition function in terms of twist fields in the 2-fold symmetric
product of the ``fake" CFT, analogously to \WittenKT. In a general
2-fold symmetric product CFT, the twist field ${\cal E}$ has OPE
with itself of the form \eqn\eeope{\eqalign{ & {\cal E}(x/2) {\cal
E}(-x/2) = x^{-3k}\Psi_x(0) + ({\rm primaries~and~ their~
descendants}),\cr & \Psi_x = \sum_{i,j} x^{\Delta_i+\Delta_j}c_{ij}
{\cal A}_i^+ {\cal A}_j^-, } } where the ${\cal A}_i^+$ and ${\cal
A}_i^-$ are Virasoro descendants of the identity (of dimension
$\Delta_i$) in the two copies of the CFT, and $c_{ij}$ are constant
coefficients. The genus two Riemann surface is represented as the
hyperelliptic curve $y^2 = \prod_{i=1}^6 (x-e_i)$. An explicit
formula for $\Psi_x$ up to order ${\cal O}(x^8)$ is given in
\GaiottoXH. We shall choose the branch cuts connecting $(e_1,e_2)$,
$(e_3,e_4)$, $(e_5,e_6)$, to correspond to the three ``filled"
handles, respectively. The fake partition function is given by the
six-point function of ${\cal E}(e_i)$'s, but dropping the
contributions from primaries in the ${\cal E}{\cal E}$ OPE \eeope\
of nearby branch points, namely the OPEs ${\cal E}(e_1){\cal
E}(e_2)$, ${\cal E}(e_3){\cal E}(e_4)$ and ${\cal E}(e_5){\cal
E}(e_6)$. Note that $x^{-3k}\Psi_{x}(0)$ transforms in the same way
as ${\cal E}(x/2) {\cal E}(-x/2)$ under conformal transformations.
Explicitly, we can write an $SL(2,{\bf C})$ covariant expression
$\tilde Z_{fake}$, \eqn\zfakeyet{ \tilde
Z_{fake}(e_1,e_2,\cdots,e_6) = \prod_{1\leq i<j\leq 6} e_{ij}^{k}
{\langle \Psi_{e_{12}}({e_1+e_2\over 2})\Psi_{e_{34}}({e_3+e_4\over
2}) \Psi_{e_{56}}({e_5+e_6\over 2}) \rangle \over
(e_{12}e_{34}e_{56})^{3k} } } Under the $SL(2,{\bf C})$ action on
the $e_i$'s, $e_i\to (ae_i+b)/(ce_i+d)$, $\tilde Z_{fake}$
transforms covariantly with weight $-2k$: \eqn\sltc{ \tilde Z_{fake}
\to \tilde Z_{fake} \prod_{i=1}^6 (ce_i+d)^{-2k} } The period matrix
$\Omega$ is determined by the $e_i$'s up to the overall $SL(2,{\bf
C})$ action. They can be mapped to the generating Siegel modular
forms via the formulae in \Igusa, as recalled in \GaiottoXH. The
fake partition function $Z_{fake}(k,\Omega)$ as a function of the
periods can be recovered from $\tilde Z_{fake}$ as \eqn\zfakerec{
Z_{fake}(k,\Omega) \propto \left[\int_{\alpha_1}{dx\over y}
\int_{\alpha_2} {xdx\over y}-(\alpha_1\leftrightarrow\alpha_2)
\right]^{2k}\tilde Z_{fake}(e_1,\cdots,e_6) } where $\alpha_1$ and
$\alpha_2$ are a pair of basis $A$-cycles. Note that the RHS of
\zfakerec\ is $SL(2,{\bf C})$ invariant. The modular group
$Sp(4,{\bf Z})$ acts as monodromies that permute the $e_i$'s. Note
that \zfakeyet\ involves an infinite series of rational functions of
the $e_i$'s, and we expect $Z_{fake}$ to have branch cuts in the
$e_i$'s; it should be invariant under $\Gamma_\infty$ only. One can
use either \fakecfp\ or \zfakeyet\ to compute $Z_{fake}(k,\Omega)$
explicitly, order by order. We will use the former in the explicit
comparison of $Z_{fake}$ with $S_0, S_1$ below.

In the cases $k=1,2,3$, by the construction of \GaiottoXH, the polar
terms of $Z_{fake}(k,\Omega)$ automatically agrees with those of
$Z_k^{mod}(\Omega)$. So the checks in the previous subsection amount
to the statement that the polar terms of $Z_{saddle}(k,\Omega)$
agrees with those of $Z_{fake}(k,\Omega)$. We will show in the next
subsection that our conjectured formula for $S_0$ and $S_1$ in
$Z_{saddle}$, remarkably, agrees with $Z_{fake}$ up to order $\nu^4$
in the $\nu\to 0$ limit, for all $k$. This lead us to conjecture
that $Z_{fake}(k,\Omega)=Z_{saddle}(k,\Omega)$. In particular,
\szero\ and \soneli\ are then predictions for the leading terms in
the large $k$ limit of $Z_{fake}(k,\Omega)$!

Finally, we conjecture the following formula for the contribution
from all handlebody geometries in the pure three-dimensional
gravity, \eqn\exactpf{ \sum_{\gamma\in \Gamma_\infty\backslash
Sp(4,{\bf Z})} \det(C\Omega+D)^{-2k} Z_{fake}(k,\gamma\cdot\Omega) =
Z_{h.b.}(k,\Omega) } And $Z_{h.b.}(k,\Omega)$ appears to capture at
least all the polar terms in the partition function
$Z_k^{mod}(\Omega)$ of the dual ECFT, if the latter exists. A few
comments are in order. Firstly, in the expansion of $Z_{fake}$ in
terms of (fake) torus one-point functions \fakecfp, terms that are
polar and constant in both $q_1$ {\sl and} $q_2$ in $Z_{fake}$ must
agree with $Z_k^{mod}$, as explained in section 2. In other words,
$Z_k^{mod} = Z_{fake} + {\cal O}(q_1,q_2)$. In appendix A.2, we show
that with certain assumptions on the regularity of
$Z_{fake}(k,\Omega)$ (away from its obvious singularities), the LHS
of \exactpf\ agrees with $Z_{fake}$ itself at least up to terms of
order ${\cal O}(q_1^0,q_2^0)$. This is consistent with the expected
property of the dual ECFT.

Another check of our proposal is in the limit where three handles of
the Riemann surface are pinched, and $Z_{fake}$ can be expanded as
\zfakeyet, with $e_{12}, e_{34}, e_{56}\to 0$. A subleading term in
the Poincar\'e series \exactpf, i.e. one with
$\gamma\not\in\Gamma_\infty$, is given by \zfakeyet\ with certain
monodromies on the $e_i$'s (we expect that there are branch cuts).
After such a permutation, at most one of $e_{12}, e_{34}, e_{56}$
may approach zero in the pinching limit. It is conceivable that the
factor $\langle \Psi\Psi\Psi \rangle$ in \zfakeyet\ does not give
extra singularities in this limit. In the representation \altfake,
we then conclude that the terms in Poincar\'e series of $Z_{fake}$
that are polar in a {\sl pair} of $\tilde q_i$'s must agree with
those in $Z_{fake}$. This is consistent with what we expect from
$Z_k^{mod}$, since terms that are polar in a pair of $\tilde q_i$'s
come from sewing together sphere three-point functions involving at
least two Virasoro descendants of the identity, and hence must agree
with $Z_{fake}$.

A priori, one expects $Z_k^{mod}$ to receive contributions from
non-handlebody geometries as well, in addition to the handlebody
contribution $Z_{h.b.}$. However, we see above that $Z_{h.b.}$
already gives the correct polar part of $Z_k^{mod}$. It is then
plausible that the non-handlebody geometries only contribute to the
non-polar part of $Z_k^{mod}$. Curiously, when $k=1$, the genus two
partition function has weight 2, and there are no weight 2 entire
Siegel modular forms (i.e. with no singularities). This means that
$Z_{h.b.}(1,\Omega)$ as given by \exactpf\ is exactly equal to
$Z_{k=1}^{mod}(\Omega)$. It would be interesting to understand why
the non-handlebody contributions would cancel in this case, and if
such cancelation could happen for other values of $k$ as well.


\subsec{Comparing the factorization of $Z_{fake}$ with $Z_{saddle}$}

Let us compute $Z_{fake}$ near the separating degeneration, up to
order $\nu^4$. It is given by \eqn\zfkvi{\eqalign{
Z_{fake}(k,\Omega) &= G(\Omega)^k \epsilon^{-2k} \left[
Z_{Vir}(\tau_1) Z_{Vir}(\tau_2) + {\epsilon^2\over 12k} {1\over 2\pi
i}\partial_{\tau_1} Z_{Vir}(\tau_1) {1\over 2\pi i}\partial_{\tau_1}
Z_{Vir}(\tau_1) \right.\cr &\left. + {5\epsilon^4\over
24k(60k+11)}\langle T*T \rangle'_{\tau_1}\langle T*T
\rangle'_{\tau_2} + {\cal O}(\epsilon^6) \right] } } where the
relation between $\tau_1,\tau_2,\epsilon$ and $\rho,\sigma,\nu$, as
well as the function $G(\Omega)$, are described in \Tuite, and
recalled in Appendix D. $Z_{Vir}(\tau) = q^{-k}\prod_{n=2}^\infty
(1-q^n)^{-1}$ is the Virasoro character.
$\langle\cdots\rangle'_\tau$ stands for the torus one-point function
in the fake CFT. $T*T(0)\equiv {\rm Res}_{z\to 0}{1\over z}T(z)T(0)$
is a dimension 4 Virasoro descendant. Its ``fake" torus one-point
function is given by (see appendix D for a derivation)
\eqn\toruone{ \langle T*T \rangle'_\tau =
\left[({1\over 2\pi i}\partial_\tau)^2-{1\over 6}{1\over 2\pi
i}\partial_\tau +{k\over 60}\right]Z_{Vir}(\tau) +
2\sum_{n=1}^\infty {\rm Tr}_{Vir(k)} L_{-n}L_n e^{2\pi i\tau(L_0-k)}
} By matching \zfkvi\ with \eqn\zsad{ Z_{saddle} =
\exp\left[{kS_0+S_1+{1\over k}S_2} + {1\over k^2}S_3+\cdots\right],
}  we find precise agreement with the expression for $S_0$, $S_1$ as
given by \szero,\soneli\ up to order $\nu^4$. Further, we can derive
all $S_l$'s from $Z_{fake}$ up to order $\nu^4$ as well, at least as
an expansion in $q,s$. We can give the closed form expressions up to order $\nu^2$,
\eqn\ssonetwo{
\eqalign{ &S_1 = -\sum_{n=2}^\infty \ln \left[(1-q^n)(1-s^n)\right]
+ (2\pi i\nu)^2\left({2q\over 1-q}\hat E_2^\sigma + {2s\over
1-s}\hat E_2^\rho-4\hat E_2^\rho \hat E_2^\sigma\right)
 + {\cal O}(\nu^4),\cr
& S_2 = {(2\pi i\nu)^2\over 12} \left({q\over 1-q}-\hat E_2^\rho
\right) \left({s\over 1-s}-\hat E_2^\sigma \right)+{\cal O}(\nu^4),
\cr & S_3 = {\cal O}(\nu^4). } } This precisely agrees with the
${\cal O}(\nu^2)$ term in \sitlee.

\newsec{Higher genera}

Our proposals have straightforward generalizations to partition functions of
higher genus $g$. The genus two modular group $Sp(4,{\bf Z})$ will be replaced by
$Sp(2g,{\bf Z})$, and the partition function transforms under the modular group with
a certain weight. Due to the conformal anomaly, there is some ambiguity in defining
the partition function. In the genus two case, it is natural to demand the partition function to have
weight $2k$, so that in the separating degeneration limit the partition function behaves like
$\epsilon^{-2k}$. For $g>2$, requiring this kind of behavior in the factorization limit
would mean that the partition function should have weight $(4-{4\over g})k$, as explained in
appendix E, by examining the case of hyperelliptic curves. To be consistent with
our genus two notation, we will continue to call this partition function
$Z_{k,g}^{mod}(\Omega)$. A more natural and convenient choice is to simply
eliminate the singularities, so that the partition function has weight $12k$. This is the equivalent to the
statement that the partition function is a holomorphic section of the $12k$-th tensor power of
the determinant line bundle ${\cal L}$ over the moduli space of the Riemann surface
\WittenKT. We will denote this partition function
by $T_{k,g}(\Omega)$, consistent with our notation in \offs.

The extension of the definition of $Z_{fake}$ to higher genera is in
principle straightforward: the genus $g$ partition function of a
general CFT can be obtained by sewing sphere $2g$-point functions,
with $g$ propagators connecting pairs of operators, corresponding to
the $g$ pinching handles; $Z_{fake}$ is defined by choosing the
1-cycles around each of the $g$ handles to be contractible in the
handlebody filling $\Sigma_g$, and sewing together all sphere
$2g$-point functions of Virasoro descendants of the identity. When
the Riemann surface is hyperelliptic, of the form
$y^2=\prod_{i=1}^{2g+2}(x-e_i)$, $Z_{fake}$ can be computed
explicitly from \eqn\zfakegen{ \tilde Z_{fake}(e_1,\cdots,e_{2g+2})
= \left(\prod_{1\leq i<j\leq 2g+2} e_{ij}\right)^{3k} {\left\langle
\prod_{s=1}^{g+1} \Psi_{e_{2s-1,2s}}({e_{2s-1}+e_{2s}\over 2})
\right\rangle\over \left(\prod_{r=1}^{g+1} e_{2r-1,2r}\right)^{3k} }
} where $e_{ij}\equiv e_i-e_j$. Note that we have chosen the
exponent $3k$ instead of $k$ in the first factor on the RHS of
\zfakegen, anticipating the full partition function
($T_{k,g}(\Omega)$) to be a modular form of weight $12k$. The
$e_i$'s are ordered so that the branch cuts connecting $e_{2s-1}$
and $e_{2s}$ ($s=1,\cdots,g$) correspond to the $g$ handles that are
filled in by the gravitational instanton. For more general Riemann
surfaces, one needs to know the mapping between the pinching
parameters and the periods, as well as the universal ``holomorphic
correction" factor, in order to calculate $Z_{fake}$ explicitly.

An obvious conjecture is that $Z_{fake}=Z_{saddle}=
\exp(k \hat S_0+\sum_{l\geq 1} k^{1-l}S_l)$ should hold for all $k$ and all genus $g$.
We used the notation $\hat S_0$ to indicate the convention that the partition function
is a weight $12k$ modular form, with no singularities.
We propose an exact expression for $\hat S_0$ to be
\eqn\szerogen{ e^{\hat S_0} = {{\cal F}(\Omega)^{12}}, } where ${\cal F}(\Omega)$ is
given by \dotn. Note that in our convention, the genus $g$ partition function $T_{\phi,g}(\Omega)$ of a chiral boson
is simply 1. This is explained in appendix E when the Riemann surface is
a hyperelliptic curve. More generally, we can regard $T_{\phi,g}(\Omega)$
as a canonical section of the line bundle ${\cal L}^{1\over 2}$. \szerogen\ is consistent
with the genus one answer and our proposal in the genus two case: $e^{\hat S_0}
=e^{S_0}\eta^{24}$ for $g=1$ and $e^{\hat S_0}
=e^{S_0}\chi_{10}$ for $g=2$.
The conjectured expression
for $S_1$, \soneli, clearly generalizes to higher genera as well.

The formula \exactpf\ for the full handlebody contribution to the
partition function, by summing over modular images of $Z_{fake}$,
also admits a straightforward generalization to $g>2$, \eqn\tkt{
\hat Z_{h.b.}(k,\Omega) = \sum_{\Gamma_\infty \backslash Sp(2g,{\bf
Z})} \det(C\Omega+D)^{-12k} \hat Z_{fake}(k,\gamma\cdot\Omega), }
where the hat emphasizes our convention for the weight $12k$
partition function. In particular, $Z_{fake}$ is $\Gamma_\infty$
invariant by the tree level factorization of the correlation
functions of Virasoro descendants of the identity. A crucial new
feature in the $g>2$ case is that the period matrix of a Riemann
surface lies in a subspace of Siegel upper-half space of nonzero
codimension. The genus $g$ partition function in general will not be
a Siegel modular form (or quotients thereof), but a Teichm\"uller
modular form. Nevertheless, \tkt\ could still be well defined for
$g>2$ and give the full handlebody contribution. Notably, modular
invariance and the factorization of $Z_{grav}(k,\Omega)$ (when
handles are pinched) as given by \tkt\ appear to be manifest. The
factorization property follows from: (a) $Z_{fake}$ is singular only
when the pinching 1-cycle is contractible in the filling
three-manifold; and (b) when such a 1-cycle is pinched, $Z_{fake}$
factorizes correctly by construction. If one could also fix the
non-handlebody contributions to all genera, then one would be able
to reconstruct all correlation functions of the dual ECFT by
expanding the partition functions near various degenerating limits
of the Riemann surfaces \FriedanUA.

\newsec{Non-handlebodies}

Let us briefly describe the case when the gravitational instanton,
represented by a hyperbolic three-manifold $M$ with conformal
boundary $\Sigma_g$, is not a handlebody. In general, $M$ can be
modeled as ${\bf H}_3/G$, where $G$, being the fundamental group of
$M$, is a Kleinian group and is not a Schottky group. This would be
the case whenever $G$ is not freely generated.\foot{If $G$ is freely
generated, but not purely loxodromic, it can be ``approximated" as a
limit of Schottky groups, as subsets of $SL(2,{\bf C})$. The formula
\dotn\ then suggests that the contribution from such instantons
should vanish, as some of the $q_\gamma$'s approach 1 in the limit.}

A simple example of such a non-handlebody $M$ is as
follows.\foot{Thanks to D. Gaiotto, E. Witten, C. McMullen for
pointing out this example.} First consider a quotient of ${\bf H}_3$
by a Fuchsian group (a subgroup of $SL(2,{\bf R})\subset SL(2,{\bf
C})$), resulting in a hyperbolic three-manifold whose conformal
boundary is two copies of $\Sigma_g$. We will call it $\widetilde
M_1$. The metric on $\widetilde M_1$ can be written as \eqn\meti{
ds^2 = d\rho^2 + \cosh^2\rho ds_\Sigma^2 } where $ds_\Sigma^2$ is a
hyperbolic metric on $\Sigma_g$. Suppose $\Sigma_g$ further admits a
fixed-point free orientation reversing involution. Let $\iota$ be
this involution together with the ${\bf Z}_2$ symmetry $\rho\mapsto
-\rho$. Then the identification of $\widetilde M_1$ by $\iota$ gives
a hyperbolic three-manifold $M_1$, with conformal boundary
$\Sigma_g$, that is not a handlebody. Although the manifold $M_1$
was constructed assuming the existence of a ${\bf Z}_2$ involution
on $\Sigma_g$, a hyperbolic metric exists on $M_1$ for {\sl any} complex
structure on $\Sigma_g$. Furthermore, the conjugacy classes of the corresponding Kleinian group
$G$ varies holomorphically with the complex moduli of $\Sigma_g$.\foot{
$M_1$ belongs to an ``extreme" class of hyperbolic three-manifolds with
conformal boundary $\Sigma_g$, in that $\pi_1(\Sigma_g)$ injects into $\pi_1(M_1)$.
Most hyperbolic three-manifolds with conformal boundary $\Sigma_g$ do not have this
property; although they may be obtained (topologically) from such manifolds
with $\pi_1$-injective boundary
by attaching 2-handles.
}

While a handlebody filling $\Sigma_g$ is invariant under
$\Gamma_\infty\subset Sp(2g,{\bf Z})$, $M_1$ is invariant under a
different subgroup $Z(\iota)$ -- the commutant of $\iota$ in
$Sp(2g,{\bf Z})$. There are other modular images of $M_1$, labeled
by the coset $Z(\iota)\backslash Sp(2g,{\bf Z})$. 
For example
suppose $g=2$, and then $\iota$ is given by the $Sp(4,{\bf Z})$ matrix
\eqn\iost{ \iota = \pmatrix{ 0& 1 & 0 & 0\cr 1 & 0 & 0 & 0\cr 0 & 0
& 0 & 1\cr 0 & 0 &1 &0 } } The period matrix $\Omega$ should then
satisfy $\Omega = -\overline{\iota\cdot \Omega}$, namely $\rho =
-\overline{\sigma}$, $\nu\in i{\bf R}$. 

We have not computed the regularized Einstein-Hilbert action of $M_1$, or any
other non-handlebody instantons. 
It is also unclear to us how to generalize our conjectured formula the fake CFT partition
function $Z_{fake}(k,\Omega)$ to the case when $M$ is not a
handlebody.

\newsec{Summary and questions}

We have provided nontrivial evidences for the conjectures in section
2. Our most conservative conjecture is the statement $\ln
Z_{fake}(k,\Omega)=kS_0+S_1+{\cal O}(1/k)$, both sides being well
defined and explicitly computable as a series expansion in the
pinching parameter $\epsilon$. We have checked it up to order
$\epsilon^4$. It would be nice to prove it. It would also be nice to
compute $1/k$ corrections in $Z_{saddle}$ as loop corrections in the
chiral $SL(2,{\bf R})$ WZW model.

One of the most important questions is how to compute the
contributions from the non-handlebody gravitational instantons to
the partition function. We have seen that the contribution from
handlebodies alone already captures the polar part of the expected
dual ECFT partition function. This suggests that the
non-handlebodies may only contribute to the non-polar part of the
partition function. 


On the CFT side, the most important question is whether the dual
ECFTs actually exist as unitary CFTs. If we can sum up all the
gravitational instanton contributions, we should obtain the full
partition function of the dual ECFT. Such partition functions are
modular invariant, and should factorize correctly in the
degenerating limits of the Riemann surface; the contribution from
the handlebodies already seems to satisfy these consistency
conditions by itself. According to \FriedanUA, all correlation
functions of the CFT can then be consistency recovered from these
partition functions. Even if this is the case, the resulting CFT is
not obviously unitary. These questions are left to future work.

\bigskip

\centerline{\bf Acknowledgement}
I'm indebted to S. Minwalla for raising the questions that motivated this work,
and for collaborations through a large part of it. I'm grateful to D. Gaiotto,
D. Jafferis, A. Strominger and E. Witten for comments on earlier drafts of the paper, and to
C. McMullen for explaining to me aspects of hyperbolic three-manifolds.
I would like to thank C. Beasley, V. Bouchard, F. Denef, J. Maldacena,
A. Maloney, G. Moore, B. Pioline, A. Simons, M. Spradlin,
A. Tomasiello, A. Volovich, D. Zagier for discussions at various occasions.
I also thank the hospitality of Ecole Normale Supérieure, LPTHE, and the organizers
of XXXVIIth Paris Summer Institute on {\sl Black Holes, Black Rings and
Modular Forms}, where this work was initiated.
I am
supported by a Junior Fellowship from the Harvard Society of
Fellows.

\bigskip

\appendix{A}{Genus two Siegel modular forms}

\subsec{Some basic properties and conventions}

The genus two weight $\half$ theta series are defined as a function
of the the period matrix $\Omega$ as \eqn\thetasa{ \Theta[{\bf
a},{\bf b}] = \sum_{{\bf n}\in {\bf Z}\times{\bf Z}} \exp\left[\pi
i({\bf n}+{\bf a})\cdot\Omega\cdot({\bf n}+{\bf a}) + 2\pi i ({\bf
n}+{\bf a})\cdot {\bf b} \right] } where ${\bf a},{\bf b}$ take
values among the vectors $(0,0),(0,\half),(\half,0),(\half,\half)$.
We will also use an alternative notation, denoting $\Theta[(\half
a_1,\half a_2),(\half b_1,\half b_2)]$ by $\Theta_{a_1a_2b_1b_2}$.
Only 10 out of 16 expressions in \thetasa\ are nonvanishing. They
are \eqn\thenon{
\Theta_{0000},~~\Theta_{0001},~~\Theta_{0010},~~\Theta_{0011},~~\Theta_{0100},~~
\Theta_{0110},~~\Theta_{1000},~~\Theta_{1001},~~\Theta_{1100},~~\Theta_{1111}.
} A set of generators of the ring of entire Siegel modular forms
(according to the convention of \Tuite) are \eqn\gennsig{\eqalign{ &
\tilde\psi_4 = {1\over 4}\sum_{{\bf a},{\bf b}} \Theta[{\bf a},{\bf
b}]^8, \cr &\Delta_{10} ={1\over 2^{12}}\prod_{{\bf a},{\bf b}}
\Theta[{\bf a},{\bf b}]^2, \cr & F_{12} = {1\over 4}\sum_{{\bf
a},{\bf b}} \Theta[{\bf a},{\bf b}]^{24}, \cr &\tilde\psi_6 =
P_0^3-9P_0(P_1^2+P_2^2+P_3^2-4P_4^2)+54 P_1 P_2 P_3, }} where the
$P_i$ are related to the $\Theta$'s by \eqn\poth{ \eqalign{&
\Theta_{0000}^4 = P_0+P_1+P_2+P_3,\cr & \Theta_{0001}^4 =
P_0-P_1+P_2-P_3,\cr & \Theta_{0010}^4 = P_0+P_1-P_2-P_3,\cr &
\Theta_{0011}^4 = P_0-P_1-P_2+P_3,\cr & \Theta_{0100}^4 =
2P_1+2P_4,\cr &\Theta_{0110}^4 = 2P_1-2P_4,\cr &\Theta_{1000}^4 =
2P_2+2P_4,\cr &\Theta_{1001}^4 = 2P_2-2P_4,\cr &\Theta_{1100}^4 =
2P_3+2P_4,\cr &\Theta_{1111}^4 = 2P_3-2P_4. } } Note that there are
5 linear relations among the 10 $\Theta^4$'s, so that \poth\
consistently determine $P_{0,1,2,3,4}$. The generators
$\psi_4,\psi_6,\chi_{10},\chi_{12}$ of \GaiottoXH\ are related to
those in \gennsig\ by \eqn\rels{\eqalign{ &\psi_4 = {1\over
4}\tilde\psi_4, \cr &\psi_6 = {1\over 16}\tilde\psi_6, \cr
&\chi_{10} = 4\Delta_{10}, \cr & \chi_{12} = {96\over
2^{13}3^4}({9\over 11}F_{12}-\tilde\psi_4^3+{2\over
11}\tilde\psi_6^2). } }

\subsec{Averaging over the modular group}

To begin let us recall the case of $SL(2,{\bf Z})$ modular forms.
See \DijkgraafFQ\ for example. Given a $\Gamma_\infty$-invariant
function $h(\tau)$, its Poincar\'e series of weight $w$ is defined
by the summation over $SL(2,{\bf Z})$ images, \eqn\modimag{ {\bf
P}_w h (\tau)= \sum_{\gamma\in \Gamma_\infty\backslash SL(2,{\bf Z})}
(c\tau+d)^{-w} h\left({a\tau+b\over c\tau+d} \right) } For $w\geq
4$, it is well known that ${\bf P}_w (q^n)$ $(n>0)$ span the space
of weight $w$ cusp forms. In particular, they vanish when
$w=4,6,8,10,14$. For these values of $w$, a weight $w$ weakly
holomorphic modular form $f_w(\tau)=\sum a_n q^n$ is given by the
Poincare series of its polar (and constant) part, \eqn\polss{
f_w^-(\tau) = \sum_{n\leq 0} a_n q^n. } Namely, $f_w(\tau) = {\bf
P}_w f_w^-(\tau)$. The situation is more complicated for higher
weights. When there are weight $w$ cusp forms, ${\bf P}_w q^n$ can
be nonzero for arbitrarily large $n$. They can be determined in
terms of the cusp forms as follows. Define the inner product
\eqn\inner{ \langle f,g\rangle_w = \int_{\cal F} f(\tau)
\overline{g(\tau)} \tau_2^{w-2}d^2\tau, } where ${\cal F}$ is the
fundamental domain for $SL(2,{\bf Z})$. If $\chi(\tau)=\sum_{n>0}
b_n q^n$ is a cusp form, then \eqn\innera{ \langle \chi, {\bf P}_w
q^n\rangle_w = (w-2)! (4\pi n)^{1-w} b_n. } Let $\chi_i$ be a basis
of weight $w$ cusp forms, with $\langle \chi_i,\chi_j
\rangle_w=c_{ij}$, then we have (for $n>0$) \eqn\psn{ {\bf P}_w q^n
= \sum_{i,j} \langle \chi_i, {\bf P}_w q^n\rangle_w
(c^{-1})^{ij}\chi_j(\tau) } Things are simple if we only want to
compare the polar terms. For positive weight $w$, if $h(\tau)$ is a
$\Gamma_\infty$-invariant function, that is regular away from
$\tau=i\infty$, and if $f_w(\tau)={\bf P}_w h(\tau)$, then
$f_w^-(\tau) = h^-(\tau)$. We can also relax the positivity
condition on $w$, while assuming that ${\bf P}_w h$ converges
absolutely and is regular away from $\tau=i\infty$, then ${\bf P}_w
h(\tau) = h(\tau) + {\cal O}(q^0)$.

Now let us turn to $Sp(4,{\bf Z})$ Siegel modular forms. Given a
$\Gamma_\infty$-invariant function $h(\Omega)$, its $Sp(4,{\bf Z})$
Poincar\'e series of weight $w$ is defined as \eqn\avedsie{ {\bf
P}_w h(\Omega)= \sum_{\gamma\in \Gamma_\infty\backslash Sp(4,{\bf
Z})} (\det(C\Omega+D))^{-w} h(\gamma\cdot\Omega) } The Fourier basis
is \eqn\ett{ e_T(\Omega) = e^{2\pi i{\rm Tr}(T\Omega)}, } where $T$
is a symmetric $2\times 2$ matrix with integer entries. $e_T$ is
invariant under $\Gamma_\infty'\subset Sp(4,{\bf Z})$, consisting of
elements of the form \eqn\eless{ \pmatrix{\pm {\bf 1} & B\cr 0 &
\pm{\bf 1} } } The sum \eqn\sumet{ h_T(\Omega)=\sum_{\gamma\in
\Gamma_\infty'\backslash \Gamma_\infty} e_T(\gamma\cdot \Omega) =
\sum_{A\in PSL(2,{\bf Z})} e_{A^T T A}(\Omega) } converges for
positive definite $T$, with $\Omega$ taking values on the Siegel
upper half space. These are the analogs of the $q^n$'s ($n>0$) in
the $SL(2,{\bf Z})$ case.

For a pair of cusp forms $f(\Omega), g(\Omega)$ of weight $w$, one
can define their inner product \eqn\inensig{ \langle f, g\rangle_w =
\int_{\cal F} f_w(\Omega)\overline{g(\Omega)} (\det {\rm
Im}\Omega)^{w-3} d\Omega d\overline{\Omega}, } where we wrote
$d\Omega\equiv d\rho d\nu d\sigma$, and ${\cal F}$ for the
fundamental domain of $Sp(4,{\bf Z})$ on the Siegel upper half space
${\bf H}$. If $f(\Omega)=\sum_{T>0} a_T e_T(\Omega)$ is a cusp form,
then \eqn\cupsfs{ \langle f, {\bf P}_wh_T\rangle_w = a_T
\int_{M_{2\times 2}\geq 0} e^{-2\pi Tr(TM)}(\det M)^{w-3}dM } We see
that once again, ${\bf P}_wh_T$ are generically nonvanishing
($T>0$), and can be expressed as a linear combination of basis cusp
forms using \cupsfs.

In practice computing ${\bf P}_w h(\Omega)$ based on \cupsfs\ is not
easy. Things simplify if we only look at the polar terms. One should
be cautious that the moduli space of a genus two Riemann surface is
not the quotient of ${\bf H}$ by $Sp(4,{\bf Z})$, but rather the
quotient of ${\bf H}-{\cal D}$, where ${\cal D}$ is the set of
$Sp(4,{\bf Z})$ images of $\nu=0$. We are interested in Siegel
modular forms of the form $f_w(\Omega) =
T(\Omega)\chi_{10}(\Omega)^{-k}$, where $T(\Omega)$ is a (weight
$w+10k$) entire modular form. Near $q=s=\nu=0$, we have $f_w(\Omega)
= {\cal O}(q^{-k}s^{-k}\nu^{-2k})$. A useful fact is the following
simple lemma.

{\bf Lemma 1.} If $w<10$, $f_w$ is a weight $w$ $Sp(4,{\bf Z})$
modular forms possibly with poles along ${\cal D}$, and if
$f_w(\Omega) = {\cal O}(q,s)$, then $f_w(\Omega)=0$.

To see this, suppose $f_w=T\chi_{10}^{-k}$, and
$T=P(\psi_4,\psi_6,\chi_{12})+\chi_{10} Q(\psi_4,\cdots)$, where $P$
is a nonzero polynomial in $\psi_4,\psi_6,\chi_{12}$. Restricting to
$\nu=0$, we have $P(E_4(\rho)E_4(\sigma),
E_6(\rho)E_6(\sigma),\Delta_{12}(\rho)\Delta_{12}(\sigma))={\cal
O}(q^{k+1},s^{k+1})$. This is only possible if
$P=\Delta_{12}(\rho)^{k+1}(\cdots)+\Delta_{12}(\sigma)^{k+1}(\cdots)$,
which would require $w+10k\geq 12(k+1)$, contradicting the
assumption $w<10$.

 Generally,
suppose $h(\Omega)$ is a $\Gamma_\infty$-invariant holomorphic
function, possibly with finite order poles along $\rho= i\infty$,
$\sigma= i\infty$, or $\nu= 0$, as well as their $\Gamma_\infty$
images, but it is otherwise regular everywhere on $\overline{\bf
H}$. Suppose that the Poincar\'e series ${\bf P}_w h(\Omega)$
converges absolutely. Then we have

{\bf Lemma 2.} ${\bf P}_w h(\Omega) = h(\Omega) + {\cal
O}(q^0,s^0).$

We need to show that for $\gamma\in Sp(4,{\bf Z})-\Gamma_\infty$,
$h(\gamma\cdot\Omega) = {\cal O}(q^0) + {\cal O}(s^0)$. Suppose
${\bf P}_w h(\Omega)=T\chi_{10}^{-k}$ for some entire Siegel modular
form $T$. Writing
$$
\gamma = \pmatrix{A & B\cr C & D},~~~~C\not=0,
$$
we can consider two cases: (a) $\det C\not=0$. In this case,
$\gamma\cdot \Omega = AC^{-1}+{\cal O}({\nu\over\rho\sigma})$ as
$\rho,\sigma\to i\infty$, and we have
$$
\left| h(\gamma\cdot\Omega) \right| < N_\gamma \left|{
\det(C\Omega+D)\over\nu}\right|^{2k}
$$
for small $\nu$. (b) ${\rm rk}C=1$. In this case, we have
$$
\gamma\cdot\Omega \sim {\rho\sigma AC^*\over \det (C\Omega+D)}
$$
where $C^*$ stands for the cofactor matrix of $C$. Then we can bound
$$
\left| h(\gamma\cdot\Omega) \right| < N_\gamma \left|e^{-2\pi
ik\parallel AC^*\parallel {\rho\sigma\over
\det(C\Omega+D)}}\right|\left|{ \det(C\Omega+D)\over\nu}\right|^{2k}
$$
for small $\nu$. In either case, the image of $h(\Omega)$ under
$\gamma$ in the Poincar\'e series can only contribute to terms of
order ${\cal O}(q^0,s^0)$, and never to terms that that polar in
{\sl both} $q$ and $s$, hence proving the lemma.

In other words, for a function $f(\Omega)=\sum a_{n,m,r} q^n s^m \sin(\pi\nu)^{2r}$, if we define
its polar part as
\eqn\polsieg{ f^-(\Omega) = \sum_{n,m<0,\,r\in{\bf Z}} a_{n,m,r} q^n s^m \sin(\pi\nu)^{2r}, }
then $({\bf P}_w h)^-=h^-$ for $h(\Omega)$ satisfying the regularity condition above. Note that
despite the summation in \polsieg\ is over all $r$, $a_{n,m,r}$ is only non-vanishing
for a finite set of values of $r$ for the type of modular forms we are considering.
One should also be cautious that our definition of polar part is {\sl not} $\Gamma_\infty$ invariant.

In our application, the handlebody contribution to the partition
function $Z_{h.b.}(k,\Omega)$ is given by the Poincar\'e series of
$Z_{fake}(k,\Omega)$, and if we assume that $Z_{fake}$ satisfy the
regularity criteria of $h(\Omega)$ above,\foot{This is not quite the
case in general: the genus one answer,
$Z_{fake}^{g=1}=q^{-k}\prod_{n=2}^\infty (1-q^n)^{-1} =
q^{-k+{1\over 24}}(1-q)\eta(\tau)^{-1}$, diverges at the $SL(2,{\bf
Z})$ images of $\tau=i\infty$ as well. Nevertheless, this problem
can be fixed if we multiply $Z_{fake}^{g=1}$ by $\eta(\tau)$, while
raising the weight of the Poincar\'e series by $\half$. In the genus
two case, it is likely that multiplying $Z_{fake}$ by
$\chi_{10}^{1\over 24}$ will suffice.} then the polar terms of
$Z_{h.b.}(\Omega)$ and $Z_{fake}(k,\Omega)$ in $q,s$ must agree. On
the other hand, the polar terms in $q,s$ of $Z_k^{mod}$ do not
involve the contribution from three-point functions of nontrivial
primaries, and hence necessarily agrees with $Z_{fake}$ (if a
consistent ECFT partition function exists). Therefore we conclude
that the handlebody contribution already gives all the correct polar
terms of $Z_k^{mod}$. In fact, the terms of order $q^0s^{n\leq 0}$
and $q^{n\leq 0}s^0$ in $Z_{fake}$ also agree with $Z_k^{mod}$, as
they do not involve nontrivial primaries either. One might then
expect $Z_{h.b.}={\bf P}_{2k} Z_{fake}$ to capture these terms as
well. This does not follow from our lemma, although there could be
better estimates on ${\bf P}_w h(\Omega)-h(\Omega)$.

\appendix{B}{The Schottky parameterization}

\subsec{Generalities}

In this subsection we describe some useful properties of the
Schottky parameterization of a Riemann surface. The Schottky group
$\Gamma$ is a subgroup of $SL(2,{\bf C})$ freely generated by $g$
elements $\gamma_1,\cdots\gamma_g$, which acts on ${\bf P}^1$ by
Mobius transformation. It is convenient to parameterize a group
element $\gamma$ by its fixed points $\xi,\eta$ and its multiplier
$q_\gamma$, \eqn\abcc{ \eqalign{ {\gamma(z)-\eta\over \gamma(z)-\xi}
= q_\gamma {z-\eta\over z-\xi}. } } Geometrically, $\gamma$ maps a
circle $C$ around $\xi$ to another circle $C'$ around $\eta$; it
maps the domain outside of $C$ (or $C'$) to the disc bounded by $C'$
(or $C$).

A Riemann surface $\Sigma_g$ of a given complex structure can be
realized as the quotient of ${\bf P}^1$ (excluding a suitable zero
measure set) by $\Gamma$. Specializing to the genus two case, we
will choose a pair of generators of $\Gamma$, $\alpha$ and $\beta$,
\eqn\abscab{ \eqalign{ {\alpha(z)-\eta_1\over \alpha(z)-\xi_1} =
q_\alpha {z-\eta_1\over z-\xi_1},~~~~~ {\beta(z)-\eta_2\over
\beta(z)-\xi_2} = q_\beta {z-\eta_2\over z-\xi_2}. } } The Schottky
space is parameterized by
$q_\alpha,q_\beta,\xi_1,\xi_2,\eta_1,\eta_2$, up to the $SL(2,{\bf
C})$ action by conjugation. The relation between the Schottky
parameters and the period matrix elements $\rho,\sigma,\nu$ is given
for example in \LeblancYY, by the following formulae \eqn\relach{
\eqalign{ & e^{2\pi i\rho} = q_\alpha
\prod_{\gamma=\beta^n\cdots\beta^m} \left({\gamma(\eta_1)
-\eta_1\over \gamma(\xi_1)-\eta_1} {\gamma(\xi_1)-\xi_1\over
\gamma(\eta_1)-\xi_1}\right), \cr & e^{2\pi i\sigma} = q_\beta
\prod_{\gamma=\alpha^n\cdots\alpha^m}
\left({\gamma(\eta_2)-\eta_2\over \gamma(\xi_2)-\eta_2}
{\gamma(\xi_2)-\xi_2\over \gamma(\eta_2)-\xi_2}\right), \cr &
e^{2\pi i\nu} = {\eta_{12}\xi_{12}\over (\xi_1-\eta_2)
(\eta_1-\xi_2)} \prod_{\gamma=\alpha^n\cdots\beta^m}
\left({\gamma(\eta_1)-\eta_2\over \gamma(\xi_1)-\eta_2}
{\gamma(\xi_1)-\xi_2\over \gamma(\eta_1)-\xi_2}\right). } } The
product in the first line runs through all distinct elements
$\gamma$ corresponding to a word with $\beta$ or $\beta^{-1}$ on the
left and right ends. The product in the second line is over elements
with $\alpha^{\pm1}$ on the left and right ends. The product in the
last line is over elements whose word starts with $\alpha^{\pm1}$
and ends with $\beta^{\pm1}$.

The Schottky parameterization is $\Gamma_\infty$ invariant.
$\Gamma_\infty$ as a subgroup of $Sp(4,{\bf Z})$ is generated by
integral shifts of the matrix elements of $\Omega$, as well as the
$SL(2,{\bf Z})$ transformations that acts on $\Omega$ as
$\Omega\mapsto A\Omega A^T$. The invariance under integral shifts of
$\Omega$ is clear from \relach. It is also clear that the $SL(2,{\bf
Z})$ transformation exchanging $\rho$ with $\sigma$ corresponds to
swapping the generators $\alpha,\beta$ in $\Gamma$. On the other
hand, the $SL(2,{\bf Z})$ transformation sending
$\rho\mapsto\rho+\sigma$, $\sigma\mapsto \sigma$, corresponds to
redefining the two generators of $\Gamma$ to be $\alpha$ and
$\alpha\beta$. In particular, it follows that ${\cal F}(\Omega)$
\dotn, as well as out conjectured formulae for $S_0$ and $S_1$, are
invariant under $\Gamma_\infty$.

\subsec{Schottky parameters in terms of periods up to ${\cal
O}(\nu^2)$}

The relation between the Schottky parameters and the periods
\relach\ is rather complicated. It is useful to expand it explicit
in $\nu$, in the $\nu\to 0$ limit (separating degeneration). In
terms of the Schottky parameters, this limits corresponds to
separating $\xi_1,\eta_1$ and $\xi_2,\eta_2$ at a large distance
$L$; $\nu$ scales like $1/L^2$.

To leading nontrivial order, only the product over elements
$\beta^n$ ($n\not=0$) contribute in the first line of \relach. In
fact, straightforward calculation shows that \eqn\difft{ \eqalign{ &
{\beta^n(\eta_1) -\eta_1\over \beta^n(\xi_1)-\eta_1}
{\beta^n(\xi_1)-\xi_1\over \beta^n(\eta_1)-\xi_1} =
1+{(\eta_1-\xi_1)^2(\eta_2-\xi_2)^2\over L^4 \left(q_\beta^{n\over
2}-q_\beta^{-{n\over 2}}\right)^2}+{\cal O}({1\over L^5}) } }
whereas for Schottky elements of the form
$\gamma=\beta\cdots\alpha\cdots\beta$, \eqn\sot{ {\gamma(\eta_1)
-\eta_1\over \gamma(\xi_1)-\eta_1} {\gamma(\xi_1)-\xi_1\over
\gamma(\eta_1)-\xi_1}=1+{\cal O}({1\over L^8}) } Plugging this into
the first line of \relach, we find \eqn\expanone{ \eqalign{ &
e^{2\pi i\rho} = q_\alpha \left[
1+{(\eta_1-\xi_1)^2(\eta_2-\xi_2)^2\over L^4}\sum_{n\not=0}
\left(q_\beta^{n\over 2}-q_\beta^{-{n\over 2}}\right)^{-2}+{\cal
O}({1\over L^5}) \right] \cr & =q_\alpha \left[
1+{(\eta_1-\xi_1)^2(\eta_2-\xi_2)^2\over L^4}{1-E_2(\tau_2)\over 12}
+{\cal O}({1\over L^5}) \right] } } and similarly for $e^{2\pi
i\sigma}$. Here $E_2(\tau)$ is the second Eisenstein series. For
brevity we will often use express the Einstein series in terms of
$\hat E_n^\tau$, defined below \foform. In particular,
$E_2(\tau)=1-24\hat E_2^\tau$, $E_4(\tau)=1+240\hat E_4^\tau$.

To calculate $\nu$, the first factor on the RHS of the third line of
\relach\ gives the dominant contribution, \eqn\expnu{ 2\pi i\nu =
{(\eta_1-\xi_1)(\eta_2-\xi_2)\over L^2}+{\cal O}({1\over L^3}) } The
next order corrections come from the product over $\gamma$ of the
form $\alpha^n\cdots\beta^m$ ($n,m\not=0$), of \eqn\estitw{
{\gamma(\eta_1) -\eta_2\over \gamma(\xi_1)-\eta_2}
{\gamma(\xi_1)-\xi_2\over \gamma(\eta_1)-\xi_2}=1+{\cal O}({1\over
L^6}) } \expanone\ and \expnu\ express $\rho,\sigma,\nu$ in terms of
the Schottky parameters to order $\nu^2$.

\subsec{${\cal O}(\nu^4)$}

In this subsection, we will carry out the computation in the
previous section to the next order $\nu^4$. One must expand $\nu$ in
terms of the Schottky parameters to order $1/L^6$. We will omit the
explicit expression as it is too lengthy. Inverting it, we can
express $1/L^2$ in terms of $\nu$, and express everything as an
expansion in powers of $\nu$. Define \eqn\qgama{ A_\gamma \equiv
{\gamma(\eta_1) -\eta_1\over \gamma(\xi_1)-\eta_1}
{\gamma(\xi_1)-\xi_1\over \gamma(\eta_1)-\xi_1} } as in the product
in the first line of \relach. After tedious but straightforward
calculations, we find \eqn\agmt{ \eqalign{ & A_{\beta^n} = 1+{(2\pi
i\nu)^2 +({1\over 12}-8\hat E_2^\alpha \hat E_2^\beta)(2\pi i\nu)^4
\over (q_\beta^{n\over 2}-q_\beta^{-{n\over 2}})^{2}} + {(2\pi
i\nu)^4\over (q_\beta^{n\over 2}-q_\beta^{-{n\over 2}})^4} + {\cal
O}(\nu^6) \cr & A_{\alpha^n\beta^m} = 1+{(2\pi i\nu)^3\over
(q_\alpha^{n\over 2}-q_\alpha^{-{n\over 2}})^{2}(q_\beta^{m\over
2}-q_\beta^{-{m\over 2}})^{2} } + {\cal O}(\nu^4) \cr &
A_{\beta^{n}\alpha^m\beta^{r}} = 1+{(2\pi i\nu)^4 \over
(q_\beta^{n\over 2}-q_\beta^{-{n\over 2}})^{2} (q_\alpha^{m\over
2}-q_\alpha^{-{m\over 2}})^{2} (q_\beta^{r\over 2}-q_\beta^{-{r\over
2}})^{2} } + {\cal O}(\nu^6) } } where
$q_\alpha\equiv e^{2\pi i\tau_1}, q_\beta\equiv e^{2\pi i\tau_2}$,
and $\hat E^\alpha_n\equiv \hat E_n(\tau_1)$, $\hat E^\beta_n\equiv
\hat E_n(\tau_2)$. Furthermore, we have \eqn\moreexps{ \eqalign{ &
\prod_{n\not=0}A_{\beta^n} = 1+(2\pi i\nu)^2 2\hat E_2^\beta+ (2\pi
i\nu)^4\left[ -16 \hat E_2^\alpha (\hat E_2^\beta)^2+ 2 (\hat
E_2^\beta)^2 + {1\over 6}E_4^\beta \right] + {\cal O}(\nu^6),\cr &
\prod_{n,m,r\not=0} A_{\beta^n\alpha^m\beta^r} = 1+(2\pi i\nu)^4 8
\hat E_2^\alpha (\hat E_2^\beta)^2
 + {\cal O}(\nu^6). } }
The periods $\rho,\sigma$ are then related by \eqn\qhgaa{
\eqalign{e^{2\pi i\rho} &= q_\alpha \prod_{n\not=0} A_{\beta^n}
\prod_{n,m,r\not=0} A_{\beta^n\alpha^m\beta^r} \cdot(1+{\cal
O}(\nu^6)) \cr &= q_\alpha\left\{1+(2\pi i\nu)^2 2\hat E_2^\beta +
(2\pi i\nu)^4 \left[ 2(\hat E_2^\beta)^2 -8\hat E_2^\alpha(\hat
E_2^\beta)^2 + {1\over 6}\hat E_4^\beta \right] + {\cal O}(\nu^6)
\right\} \cr e^{2\pi i\sigma}&= q_\beta\left\{1+(2\pi i\nu)^2 2\hat
E_2^\alpha + (2\pi i\nu)^4 \left[ 2(\hat E_2^\alpha)^2-8\hat
E_2^\beta(\hat E_2^\alpha)^2 + {1\over 6}\hat E_4^\alpha \right] +
{\cal O}(\nu^6) \right\} }} Using the identity \eqn\idet{ {1\over
2\pi i}\partial_\tau E_2(\tau) = {E_2(\tau)^2-E_4(\tau)\over 12}, }
we can express \eqn\stoee{ \hat E_2^\beta = \hat E_2^\sigma + (2\pi
i\nu)^2\hat E_2^\rho {-\hat E_2^\sigma+12(\hat E_2^\sigma)^2-5\hat
E_4^\sigma\over 3} +{\cal O}(\nu^4) } and similarly $\hat
E_2^\alpha$ in terms of $\hat E_2^{\rho,\sigma}$. We can then invert
\qhgaa\ and express $q_\alpha,q_\beta$ in terms of the periods,
\eqn\qabpr{ \eqalign{ & q_\alpha = e^{2\pi i\rho} \left\{ 1-(2\pi
i\nu)^2 2\hat E_2^\sigma + (2\pi i\nu)^4 \left[ 2 (\hat
E_2^\sigma)^2+{2\over 3}\hat E_2^\rho \hat
E_2^\sigma\right.\right.\cr &\left.\left.~~~~~~~~~~~ -{1\over 6}\hat
E_4^\sigma + {10\over 3}\hat E_2^\rho \hat E_4^\sigma \right] +
{\cal O}(\nu^6) \right\} \cr & q_\beta = e^{2\pi i\sigma} \left\{
1-(2\pi i\nu)^2 2\hat E_2^\rho + (2\pi i\nu)^4 \left[  2 (\hat
E_2^\rho)^2+{2\over 3}\hat E_2^\sigma \hat E_2^\rho\right.\right.\cr
&\left.\left.~~~~~~~~~~~ -{1\over 6}\hat E_4^\rho + {10\over 3}\hat
E_2^\sigma \hat E_4^\rho \right] + {\cal O}(\nu^6) \right\} } }

\appendix{C}{Expanding ${\cal F}(\Omega)$ to ${\cal O}(\nu^4)$}

To determine ${\cal F}(\Omega)$ to order $\nu^4$, we must compute
$q_{\alpha^n\beta^m}$ as well as
$q_{\alpha^{n_1}\beta^{m_1}\alpha^{n_2}\beta^{m_2}}$. After some
messy algebra, these can be expressed straightforwardly in terms of
$q_\alpha$, $q_\beta$, expanded in powers of $1/L$. Translating
$1/L^2$ to $\nu$, we find \eqn\qableng{\eqalign{ &
q_{\alpha^n\beta^m} = {(2\pi i\nu)^2\over (q_\alpha^{n\over
2}-q_\alpha^{-{n\over 2}})^2(q_\beta^{n\over 2}-q_\beta^{-{m\over
2}})^2} + {(2\pi i\nu)^3 (q_\alpha^{n\over 2}+q_\alpha^{-{n\over
2}})(q_\beta^{m\over 2}+q_\beta^{-{m\over 2}})\over
(q_\alpha^{n\over 2}-q_\alpha^{-{n\over 2}})^3(q_\beta^{m\over
2}-q_\beta^{-{m\over 2}})^3 } \cr &~~+(2\pi i\nu)^4 \left[ ({7\over
12}-8\hat E_2^\alpha \hat E_2^\beta) (q_\alpha^{n\over
2}-q_\alpha^{-{n\over 2}})^{-2}(q_\beta^{m\over 2}-q_\beta^{-{m\over
2}})^{-2} + 3 (q_\alpha^{n\over 2}-q_\alpha^{-{n\over
2}})^{-4}(q_\beta^{m\over 2}-q_\beta^{-{m\over 2}})^{-2}\right. \cr
&~~\left.+ 3 (q_\alpha^{n\over 2}-q_\alpha^{-{n\over
2}})^{-2}(q_\beta^{m\over 2}-q_\beta^{-{m\over 2}})^{-4}+14
(q_\alpha^{n\over 2}-q_\alpha^{-{n\over 2}})^{-4}(q_\beta^{m\over
2}-q_\beta^{-{m\over 2}})^{-4} \right] + {\cal O}(\nu^6),\cr &
q_{\alpha^{n_1}\beta^{m_1}\alpha^{n_2}\beta^{m_2}} = {(2\pi
i\nu)^4\over (q_\alpha^{n_1\over 2}-q_\alpha^{-{n_1\over
2}})^{2}(q_\alpha^{n_2\over 2}-q_\alpha^{-{n_2\over
2}})^{2}(q_\beta^{m_1\over 2}-q_\beta^{-{m_1\over
2}})^{2}(q_\beta^{m_2\over 2}-q_\beta^{-{m_2\over 2}})^{2}}+{\cal
O}(\nu^6). }} Taking their products, we have \eqn\prodss{ \eqalign{
& \prod_{n_1,n_2,m_1,m_2\not=0}
(1-q_{\alpha^{n_1}\beta^{m_1}\alpha^{n_2}\beta^{m_2}}) = 1-(2\pi
i\nu)^4 16(\hat E_2^\rho)^2 (\hat E_2^\sigma)^2 + {\cal O}(\nu^6)
\cr & \prod_{n,m\not=0}(1-q_{\alpha^n\beta^m}^2) = 1-(2\pi i\nu)^4
{(\hat E_4^\rho-\hat E_2^\rho)(\hat E_4^\sigma-\hat E_2^\sigma)\over
9} +{\cal O}(\nu^6) \cr & \prod_{n,m\not=0} (1-q_{\alpha^n\beta^m})
= 1-(2\pi i\nu)^2 4\hat E_2^\rho \hat E_2^\sigma + {(2\pi
i\nu)^4\over 18}\left[\hat E_2^\rho \hat E_2^\sigma + 24 (\hat
E_2^\rho)^2 \hat E_2^\sigma +24 (\hat E_2^\sigma)^2 \hat E_2^\rho
\right. \cr &~\left.+144 (\hat E_2^\rho)^2 (\hat E_2^\sigma)^2-7
\hat E_4^\rho \hat E_2^\sigma -7 \hat E_4^\sigma \hat E_2^\rho+120
\hat E_4^\rho (\hat E_2^\sigma)^2+120 \hat E_4^\sigma(\hat
E_2^\rho)^2-29 \hat E_4^\rho \hat E_4^\sigma \right] + {\cal
O}(\nu^6) } } Note that the conjugacy class represented by
$\alpha^{n_1}\beta^{m_1}\alpha^{n_2}\beta^{m_2}$ is the same if one
interchanges $(n_1,m_1)$ with $(n_2,m_2)$. Furthermore, if
$(n_1,m_1)=(n_2,m_2)$, this is not a primitive class, and should not
be included in the infinite product definition of ${\cal
F}(\Omega)$. Putting these together, \eqn\altoeg{\eqalign{&
\prod_{m=1}^\infty\prod_{\gamma\not=\alpha^{\pm 1},\beta^{\pm 1},
~prim.cl.} (1-q_\gamma^m) = 1-(2\pi i\nu)^2 4\hat E_2^\rho \hat
E_2^\sigma + {(2\pi i\nu)^4\over 3}\left[ 4 (\hat E_2^\rho)^2 \hat
E_2^\sigma +4 (\hat E_2^\sigma)^2 \hat E_2^\rho\right. \cr
&~~~~~~~\left. - \hat E_4^\rho \hat E_2^\sigma - \hat E_4^\sigma
\hat E_2^\rho +20 \hat E_4^\rho (\hat E_2^\sigma)^2+20 \hat
E_4^\sigma(\hat E_2^\rho)^2-5 \hat E_4^\rho \hat E_4^\sigma \right]
+ {\cal O}(\nu^6),\cr &\prod_{m=1}^\infty(1-q_\alpha^m) =
\prod_{m=1}^\infty (1-e^{2\pi im\rho})\times \left[1+(2\pi i\nu)^2
2\hat E_2^\rho \hat E_2^\sigma + (2\pi i\nu)^4\left(-{2\over 3}(\hat
E_2^\rho)^2 \hat E_2^\sigma
  \right.\right.
\cr &\left.\left.~~-{1\over 3}\hat E_2^\rho (\hat
E_2^\sigma)^2+6(\hat E_2^\rho)^2(\hat E_2^\sigma)^2+ {1\over 6}\hat
E_2^\rho \hat E_4^\sigma - {10\over 3}(\hat E_2^\rho)^2 \hat
E_4^\sigma -{5\over 3}\hat E_4^\rho (\hat E_2^\sigma)^2 \right)
+{\cal O}(\nu^6)\right]. }}
Finally, we arrive at the expansion for ${\cal F}(\Omega)$,
\eqn\fomehfi{ \eqalign{ & {{\cal F}(\Omega)\over \prod_{m=1}^\infty
(1-q^m)^2 (1-s^m)^2} = 1+ (2\pi i\nu)^2 4 \hat E_2^\rho \hat
E_2^\sigma +{(2\pi i\nu)^4\over 3} \left[-2 (\hat E_2^\rho)^2 \hat
E_2^\sigma - 2 \hat E_2^\rho (\hat E_2^\sigma)^2 \right. \cr &\left.
~~~~~~~~~~~~~~~~~~~~+48 (\hat E_2^\rho)^2(\hat E_2^\sigma)^2-10
(\hat E_2^\rho)^2 \hat E_4^\sigma - 10 (\hat E_2^\sigma)^2 \hat
E_4^\rho-5 \hat E_4^\rho \hat E_4^\sigma \right] + {\cal O}(\nu^6) }
}

\appendix{D}{The sewing parameters}

The holomorphic correction factor $G(\Omega)$ has an expansion in
$\epsilon$ up to order $\epsilon^4$ as \eqn\gexpe{ G(\Omega) = 1-\epsilon^2
{E_2(\tau_1) E_2(\tau_2)\over 72}
+\epsilon^4\left[{E_2(\tau_1)^2E_2(\tau_2)^2\over 6912} +
{E_4(\tau_1)E_4(\tau_2)\over 17280} \right]
+ {\cal O}(\epsilon^6). } The ${\cal O}(\epsilon^2)$ result was given in
\Tuite. The ${\cal O}(\epsilon^4)$ result is obtained by comparing the order $k$ term
in $Z_{fake}$ with our conjectured expression for $S_0$, and further verified
by comparing with the factorization of the genus two partition functions of
$k=1,2,3$ ECFTs. The sewing
parameters $\tau_1,\tau_2,\epsilon$ of \Tuite\ are related to the
period matrix elements $\rho,\sigma,\nu$ by \eqn\tuiteexp{ \eqalign{
& e^{2\pi i\rho} = e^{2\pi i\tau_1} \left\{ 1-\epsilon^2
{E_2(\tau_2)\over 12}
+\epsilon^4E_2(\tau_2)^2 \left[ {1\over 288}-{E_2(\tau_1)\over 1728}\right]
+{\cal O}(\epsilon^6) \right\}, \cr & e^{2\pi
i\sigma} = e^{2\pi i\tau_2} \left\{ 1-\epsilon^2 {E_2(\tau_1)\over
12}
+\epsilon^4E_2(\tau_1)^2 \left[ {1\over 288}-{E_2(\tau_2)\over 1728}\right]
+{\cal O}(\epsilon^6) \right\}, \cr & 2\pi i\nu = \epsilon \left[
1 + \epsilon^2 {E_2(\tau_1) E_2(\tau_2)\over 144} + {\cal
O}(\epsilon^4) \right]. } } Or inversely, \eqn\tuiteexpinv{
\eqalign{ & e^{2\pi i\tau_1} = e^{2\pi i\rho} \left\{ 1+(2\pi i\nu)^2
{E_2(\sigma)\over 12}
+(2\pi i\nu)^4 \left[ {E_2(\sigma)^2\over 288}-{E_2(\rho)E_4(\sigma)\over 1728}\right]
+{\cal O}(\nu^6) \right\}, \cr & e^{2\pi
i\tau_2} = e^{2\pi i\sigma} \left\{ 1+(2\pi i\nu)^2 {E_2(\rho)\over
12}
+(2\pi i\nu)^4 \left[ {E_2(\rho)^2\over 288}-{E_2(\sigma)E_4(\rho)\over 1728}\right]
+{\cal O}(\nu^6) \right\}, \cr & \epsilon = 2\pi i\nu \left[ 1 -
(2\pi i\nu)^2 {E_2(\rho) E_2(\sigma)\over 144} + {\cal O}(\nu^4)
\right]. } } As pointed out in \Tuite, the genus two partition
function $Z_{k,g=2}$ that naturally factorizes into one-point
functions on tori with moduli $\tau_1,\tau_2$, and sewed together
using the pinching parameter $\epsilon$, is not the modular
partition function $Z_{k,g=2}^{mod}$. Rather, it is related to
$Z_{k,g=2}^{mod}$ by \eqn\gfac{ Z_{k,g=2}(\Omega) =
{Z_{k,g=2}^{mod}(\Omega)\over G(\Omega)^k}. } Near the separating
degeneration, one has \eqn\stoteaa{\eqalign{ Z_{k,g=2}(\Omega) &=
\sum_i\epsilon^{-2k+\Delta_i} \langle {\cal A}_i \rangle_{\tau_1}
\langle {\cal A}_i\rangle_{\tau_2}, \cr &= {Z_{k,g=1}(\tau_1)
Z_{k,g=1}(\tau_2)\over \epsilon^{2k}} + {{1\over2\pi i}\partial_\tau
Z_{k,g=1}(\tau_1) {1\over2\pi i}\partial_\tau Z_{k,g=1}(\tau_2)
\over 12k\epsilon^{2k-2}} \cr &~~~+ {5\over 24k(60k+11)\epsilon^{2k-4}}
\langle T*T \rangle_{\tau_1}\langle T*T \rangle_{\tau_2}+ {\cal O}(\epsilon^{6-2k}) } } where the
second term comes from the torus one-point function of the stress
energy tensor, which is the only operator of dimension 2 in an ECFT. In general,
for a primary field ${\cal O}$ of dimension $\Delta$, the torus one-point function
$\langle {\cal O}\rangle_\tau$ is a weight $\Delta$ cusp form in an ECFT. In particular,
$\langle {\cal O}\rangle_\tau=0$ for $\Delta<12$, and $\langle {\cal O}\rangle_\tau$
can only contribute to \stoteaa\ starting at order $\epsilon^{12-2k}$. At dimension 4, the Virasoro descendants
are $\partial^2T$ and $T*T$, only the latter having a nonzero torus one-point function. To calculate
$\langle T*T\rangle_\tau$, one can first perform a conformal transformation mapping $T*T$ from
the cylinder to the complex plane ($z=e^{iw}$),
\eqn\conformalmap{ T*T= \oint_{C_2} {dw_2\over 2\pi } \oint_{C_1} {dw_1\over 2\pi i}
{T(w_1)T(w_2)\over w_1-w_2} = \oint_{C_2} {dz_2\over 2\pi iz_2}\oint_{C_1} {dz_1\over 2\pi iz_1}
{(z_1^2 T(z_1)-k) (z_2^2 T(z_2)-k)\over\ln z_1-\ln z_2} } where the contour
$C_1$ goes around $w_2$ or $z_2$. On the $z$-plane, one can compute the operator corresponding to
$T*T$, ${\cal O}_{T*T}$, by computing the integral on the RHS of \conformalmap. $C_1$ can be chosen as the sum of a contour inside $C_2$
clockwise around $z=0$, and another counterclockwise contour outside $C_2$. We end up with
\eqn\ttopera{ {\cal O}_{T*T} = (L_0-k)^2-{L_0-k\over 6}+{k\over 60}+2\sum_{n=1}^\infty L_{-n} L_n. }
\toruone\ follows from the trace of \ttopera.

\appendix{E}{Partition functions on hyperelliptic curves}

In this appendix, we recall some properties of the partition function
of a CFT with $c=24k$ on
a genus $g$ hyperelliptic curve $y^2 = \prod_{i=1}^{2g+2}(x-e_i)$, following \WittenKT.
We can write
\eqn\lsuir{ \tilde Z(e_1,\cdots,e_{2g+2})
 = \left[\prod_{1\leq i<j\leq 2g+2} (e_i-e_j)^k\right] \langle {\cal E}(e_1)\cdots {\cal E}(e_{2g+2}) \rangle, }
where ${\cal E}$ is the twist field in the 2-fold symmetric product CFT (on the $x$-plane).
Under the $SL(2,{\bf C})$ action,
\eqn\sltcc{ \eqalign{ & e_i \to {ae_i+b\over ce_i+d}, \cr
& \tilde Z \to \tilde Z \prod_{i=1}^{2g+2}(ce_i+d)^{(2-2g)k}. } }
The $e_i$'s are determined by the moduli of the hyperelliptic Riemann surface
up to the overall $SL(2,{\bf C})$ action and permutations. To construct an $SL(2,{\bf C})$
invariant expression, consider the differential
\eqn\diffsa{ \Theta={dx\over y}\wedge {xdx\over y}\wedge \cdots \wedge {x^{g-1}dx\over y}, }
regarded as a top form on the space of holomorphic 1-forms. Under $\mu$, $x$ and $y$ transform as
\eqn\xytrans{ x\to {ax+b\over cx+d},~~~~~y\to {y\over (cx+d)^{g+1} \prod_i (ce_i+d)^{1/2} }. }
As a consequence, \diffsa\ gets multiplied by
$\prod_i (ce_i+d)^{g/2}$. The
partition function, as a function of the period matrix $\Omega$ and invariant under
$SL(2,{\bf C})$, can be recovered from \lsuir,
\eqn\zztild{ Z(\Omega) \propto \left( \int_{\alpha^1\wedge\cdots\wedge \alpha^g}
\Theta\right)^{(4-{4\over g})k} \tilde Z(e_1,\cdots,e_{2g+2}) }
where the integral is understood to be on $\bigwedge^g H^1(\Sigma_g)$. The modular group $Sp(2g,{\bf Z})$
acts as monodromies on the $e_i$'s, in general permuting them; it also acts
on the basis 1-cycles, $\alpha^I, \beta_I$, $I=1,\cdots, g$.
We can choose a set of holomorphic 1-forms $\omega_I$, with
\eqn\normhol{ \int_{\alpha^I} \omega_J = \delta_J^I,~~~~\int_{\beta_I} \omega_J = \Omega_{IJ}. }
Under $Sp(2g,{\bf Z})$, they transform as
\eqn\abetra{ \eqalign{ &\alpha \to D\alpha + C\beta,~~~~~\beta \to B\alpha + A\beta, \cr
& \omega_I \to {\left((C\Omega+D)^{-1}\right)_I}^J \omega_J,~~~~\Omega \to (A\Omega+B)(C\Omega+D)^{-1}. } }
Therefore \diffsa\ transforms with
a Jacobian factor $\det(C\Omega+D)^{-1}$, and $Z(\Omega)$ transforms as an $Sp(2g,{\bf Z})$
modular form with weight $w=(4-{4\over g})k$.

For $g>2$, it is more natural to define the partition function $T(\Omega)$ to have
weight $12k$ and be free of singularities. In the hyperelliptic case, we can write
\eqn\lsuira{ \widetilde{T}(e_1,\cdots,e_{2g+2})
= \left[\prod_{1\leq i<j\leq 2g+2} (e_i-e_j)^{3k}\right]
\langle {\cal E}(e_1)\cdots {\cal E}(e_{2g+2}) \rangle, }
and similarly
\eqn\ttzz{ T(\Omega) \propto \left( \int_{\alpha^1\wedge\cdots\wedge \alpha^g}
\Theta\right)^{12k} \tilde T(e_1,\cdots,e_{2g+2}) }
$\tilde T$ has weight $-6gk$ with respect to the $SL(2,{\bf C})$ action, leading to
$T(\Omega)$ of weight $12k$ under $Sp(2g,{\bf Z})$.

A useful result is the partition function of a chiral boson $\phi$ on
the hyperelliptic curve \ZamolodchikovAE, $\tilde T_\phi(e_1,\cdots,e_{2g+2})=(\int_{\alpha^1\wedge
\cdots\wedge \alpha^g} \Theta)^{-{1\over 2}}$, and hence $T_\phi(\Omega)=1$. In the case of genus one
and two, this is equivalent to the well known results $Z_{\phi,g=1}(\tau)=\eta(\tau)^{-1}$,
$Z_{\phi,g=2}(\Omega)=\chi_{10}(\Omega)^{-{1\over 24}}$.

The fake CFT partition functions, $Z_{fake}$ (corresponding to $Z(\Omega)$), or $\hat Z_{fake}$
(corresponding to $T(\Omega)$), can be defined on a hyperelliptic Riemann surface analogously.
One arranges the $e_i$'s in pairs ($e_{2s-1},e_{2s}$), and keeps only the Virasoro descendants in
the ${\cal E}(e_{2s-1}) {\cal E}(e_{2s})$ OPE. The formulae \zztild\ and \ttzz\ generalize
to the fake partition functions as well, with the difference being that $\tilde Z_{fake}(e_1,\cdots,e_{2g+2})$
or $\tilde{\hat Z}_{fake}(e_1,\cdots,e_{2g+2})$
is not invariant under the monodromies on the $e_i$'s; in general they have branch cuts.

\listrefs

\end

****unused stuff*****

 Using the expansion for $\chi_{10}$ and for
$G(\Omega)$ (from Tuite) \eqn\chiexpe{ \eqalign{ &
\chi_{10}(\rho,\sigma,\nu) = (2\pi i\nu)^2
\Delta(\rho)\Delta(\sigma) \left[ 1 + (2\pi i\nu)^2{E_2(\rho)
E_2(\sigma)\over 12} +{\cal O}(\nu^4) \right],  } }

 Let us organize the Fourier
expansions of $f_w(\Omega)$ and $h(\Omega)$ as \eqn\fhfour{
\eqalign{ & f_w (\Omega) = \sum_{n,m,r} a_{n,m,r} q^n s^m \sin(\pi
\nu)^{2r},\cr & h(\Omega) = \sum_{n,m,r} h_{n,m,r} q^n s^m \sin(\pi
\nu)^{2r}. } } Note that if
$f_w(\Omega)=T(\Omega)\chi_{10}(\Omega)^{-k}$ where $T(\Omega)$ is
an entire Siegel modular form, then $a_{n,m,r}$ is non-vanishing
only for $$n,m,r\geq -k,~~~4(n+k)(m+k)-(r+k)^2\geq -k^2.$$ In
particular, for fixed $n,m$, $a_{n,m,r}$ is non-vanishing only for a
finite set of values of $r$. For positive weight $w$, we claim that
$f_w^-(\Omega)=h^-(\Omega)$, where the polar piece of $f_w(\Omega)$,
for instance, is defined as \eqn\polarsie{
f_w^-(\Omega)=\sum_{n,m\leq 0,\,r\in{\bf Z}} a_{n,m,r} q^n s^m
\sin(\pi\nu)^{2r} } PROVE THIS! Our definition of the polar part is
rather naive: it is not even invariant under $\Gamma_\infty$.

Conversely, if another $\Gamma_\infty$-invariant function $\tilde
h(\Omega)$ has the same polar piece $\tilde h^-(\Omega)$ as
$f_w^-(\Omega)=h^-(\Omega)$, and if $\tilde h(\Omega)$ is regular
away from the $Sp(4,{\bf Z})$ images of $\rho\to i\infty$,
$\sigma\to i\infty$ or $\nu\to 0$, then $P_w(\tilde
h(\Omega))=P_w(h(\Omega))=f_w(\Omega)$ for $w<10$. When $w\geq 10$,
$f_w(\Omega)$ is no longer determined by its polar part, and it is
more difficult to compute the Fourier coefficients of ${\bf P}_w
h(\Omega)$.

Inverting these relations, we can also write \eqn\sinv{ \eqalign{ &
q_\alpha = e^{2\pi i\rho} \left[ 1-(2\pi i\nu)^2 {1-E_2(\sigma)\over
12}+{\cal O}(\nu^3) \right], \cr & q_\beta =
 e^{2\pi i\sigma} \left[ 1-(2\pi i\nu)^2
{1-E_2(\rho)\over 12}+{\cal O}(\nu^3) \right]. } } Let us now
calculate $q_{\alpha^n\beta^m}$, for $n,m\not=0$. We find that
 \eqn\qab{\eqalign{ q_{\alpha^n\beta^m} &= {(\xi_1-\eta_1)^2(\xi_2-\eta_2)^2\over L^4}
\left( q_\alpha^{n\over 2}-q_\alpha^{-{n\over 2}} \right)^{-2}
\left( q_\beta^{m\over 2}-q_\beta^{-{m\over 2}} \right)^{-2} + {\cal
O}({1\over L^5}) \cr &=(2\pi i\nu)^2 \left( e^{\pi i n\rho}-e^{-\pi
i n\rho} \right)^{-2} \left( e^{\pi i m\sigma}-e^{-\pi i m\sigma}
\right)^{-2} + {\cal O}(\nu^3) } } The $q_\gamma$'s associated with
more general words like $\gamma=\alpha\beta\alpha\beta \cdots$ are
suppressed by more powers of $\nu$. In the end, we find the
expansion of ${\cal F}(\Omega)$, \eqn\sitpo{ {\cal F}(\Omega) =
\prod_{m=1}^\infty (1-e^{2\pi im\rho})^2(1-e^{2\pi im \sigma})^2
\left[ 1+(2\pi i\nu)^2 {1-E_2(\rho)\over 12} {1-E_2(\sigma)\over
12}+ {\cal O}(\nu^4) \right]. }  the expansion of
$(G\cdot\chi_{10})^{-1}{\cal F}(\Omega)^{12}$ in $\nu$ is then given
by \eqn\chiten{\eqalign{ {{\cal F}(\Omega)^{12}\over
G(\Omega)\chi_{10}(\Omega)} &= e^{-2\pi i(\rho+\sigma)}\left[
{1\over (2\pi i\nu)^2}+ {1-E_2(\rho)-E_2(\sigma)+{1\over
6}E_2(\rho)E_2(\sigma)\over 12} +{\cal O} (\nu^2) \right] \cr &=
{q^{-1}s^{-1}\over (2\pi i\nu)^2} -{5\over 72}q^{-1}s^{-1} +{5\over
3}(q^{-1}+s^{-1}) + 8 +{\cal O}(q,s,\nu^2) } } where $q=e^{2\pi
i\rho}, s=e^{2\pi i\sigma}$. In the $k=2$ case, we have
\eqn\ktwot{\eqalign{ \left[ {{\cal F}(\Omega)^{12}\over
G(\Omega)\chi_{10}(\Omega)}\right]^2&=e^{-4\pi i(\rho+\sigma)}
\left[ {1\over (2\pi i\nu)^4} +{1-E_2(\rho)-E_2(\sigma)+{1\over
6}E_2(\rho)E_2(\sigma)\over 6(2\pi i\nu)^2} +{\cal O}(\nu^0)
\right]\cr &= {q^{-2}s^{-2}\over (2\pi i\nu)^4} + {-{5\over
36}q^{-2}s^{-2} +{10\over 3}(q^{-2}s^{-1}+q^{-1}s^{-2}) + 16
q^{-1}s^{-1} \over (2\pi i\nu)^2}+{\cal O}(q^0,s^0,\nu^0) }}

And we will define the genus two ``partition function" of this fake CFT to be
\eqn\gluing{ Z_{fake}= \sum_{i,j} \left[q_\alpha(\xi_1-\eta_1)^2 \right]^{\Delta_i}
\left[q_\beta(\xi_2-\eta_2)^2 \right]^{\Delta_j}
\langle {\cal O}_i(\xi_1) {\cal O}_{i}(\eta_1)
{\cal O}_{j}(\xi_2) {\cal O}_{j} (\eta_2) \rangle } where ${\cal O}_i$ is a complete set of Varasoro descendants of 1,
normalized so that
\eqn\notms{ \langle {\cal O}_i(z) {\cal O}_j(0) \rangle = {\delta_{ij}\over z^{2\Delta_i}} }
$\xi_1,\eta_1$ and $\xi_2,\eta_2$ are the fixed points of the Schottky group generators $\alpha$ and $\beta$ respectively,
as before. Note that \gluing\ is $SL(2,{\bf C})$ invariant. It should also be invariant
under $\Gamma_\infty \subset Sp(4,{\bf Z})$. IS THIS TRUE??? Our conjecture is that
\eqn\zfakecon{ Z_{fake} = e^{S_1 + {1\over k}S_2 + {1\over k^2}S_3 \cdots} }
Note that the classical saddle point contribution $kS_0$ is missing from \zfakecon.
Let us check this to the leading few orders. We have
\eqn\zexp{ \eqalign{ Z_{fake} &= 1+q_\alpha^2+q_\beta^2 + q_\alpha^2q_\beta^2 (\xi_1-\eta_1)^4 (\xi_2-\eta_2)^4
{\langle T(\xi_1)T(\eta_1)T(\xi_2)T(\eta_2) \rangle\over (12k)^2} + {\cal O}(q^3,s^3) \cr
&
=1+q_\alpha^2+q_\beta^2 + q_\alpha^2q_\beta^2(1+x^2+y^2) +{1\over 3k}q_\alpha^2q_\beta^2 (x+y+xy)
+ {\cal O}(q^3,s^3)
} }
where
\eqn\xydef{ x\equiv {(\xi_1-\eta_1)^2 (\xi_2-\eta_2)^2
\over(\xi_1-\eta_2)^2 (\xi_2-\eta_1)^2 },~~~~y\equiv{(\xi_1-\eta_1)^2 (\xi_2-\eta_2)^2
\over(\xi_1-\xi_2)^2 (\eta_1-\eta_2)^2 }. }
\zexp\ precisely agrees with \expanssone\ and \sitlee\ to the leading nontrivial order. CHECK SUBLEADING ORDERS!!!

Firstly, ${\cal F}(\Omega)$ should be invariant under the shift of
matrix elements $\Omega_{mn}$ by integers. This is in fact clear
from the geometry of Schottky parameterization: the generators of
the Schottky group $\gamma_1,\cdots,\gamma_g\in SL(2,{\bf C})$ are
simply invariant under integral shifts of $\Omega_{mn}$. Secondly,
${\cal F}(\Omega)$ should be invariant under the $SL(g,{\bf Z})$
transformations which recombine the $g$ basis 1-cycles which are
contractible in the filling three-manifold. The recombination can be
realized by redefining the generators of the Schottky group and
suitably choosing the fundamental domain. The conclusion is that
${\cal F}(\Omega)$ indeed is invariant under $\Gamma_\infty$,
consistent with its role in the gravitational instanton action.

In this subsection we shall examine the geometry of the Schottky
group in the genus two case in more detail. $\Gamma$ is generated by
two elements, $\alpha$ and $\beta$. $\alpha$ maps the domain outside
a disc centered at $z_1$ to a disc centered at $\tilde z_1$, and
$\beta$ maps the domain outside a disc centered at $z_2$ to a disc
centered at $\tilde z_2$. We can write \eqn\astat{
\alpha={\rho_1\over z-z_1}+\tilde z_1,~~~~\beta={\rho_2\over
z-z_2}+\tilde z_2. } Let us denote the circle around $z_1$ by $C_1$,
and similarly $\tilde C_1, C_2$ and $\tilde C_2$. $\alpha$ maps
$C_1$ to $\tilde C_1$, while $\beta$ maps $C_2$ to $\tilde C_2$. The
fundamental domain $D$ is the area outside $C_1, C_2, \tilde C_1,
\tilde C_2$. A set of basis 1-cycles of the genus two Riemann
surface are $A_1, A_2$, which are circles winding around $z_1$ and
$z_2$, along the outside of $C_1$ and $C_2$; and $B_1, B_2$, which
are paths connected $C_1$ to $\tilde C_1$, and $C_2$ to $\tilde
C_2$.

One can pick holomorphic 1-forms $\omega_1$ and $\omega_2$, with the
property that $\oint_{A_i} \omega_j=\delta_{ij}$. Then the periods
of $\omega_i$ over the $B$-cycles give the period matrix. Let us
discuss what happens when acting on the 1-cycles by
$\Gamma_\infty\subset Sp(4,{\bf Z})$. First consider the integral
shifts of $\Omega$. If we make the shift $\rho\to \rho+1$, it
amounts to rotating $\tilde C_1$ by $2\pi$, or multiplying $\rho_1$
in \astat\ by $e^{2\pi i}$, which of course does nothing. If we
shift $\nu\to \nu+1$, it corresponds to taking $\tilde C_1$ and
moving it all the way around $\tilde C_2$. This of course does not
change the action of the Schottky group either. So we see that
$q_\gamma$'s are invariant under shifting $\Omega$ by integral
matrices.

Now let us consider the action of $SL(2,{\bf Z})\subset
\Gamma_\infty$ on $A_1, A_2$. Exchanging $A_1$ with $A_2$, of
course, simply corresponds to swapping the generators $\alpha$ and
$\beta$. Consider $A_1\to A_1+A_2$, $A_2\to A_2$. We can draw a
circle surround both $C_1$ and $C_2$, but outside $\tilde C_1$ and
$\tilde C_2$. Let us call it $C_1'$. We will also define $C_2'$ to
be the circle $\tilde C_2$. $C_1'$ is in the class $A_1+A_2$, while
$C_2'$ is in the class $A_2$. Now $\alpha(C_1')$ is a circle lying
inside $\tilde C_1$, and $\alpha\circ\beta^{-1}(C_2')$ is also a
circle lying inside $\tilde C_1$, but outside of $\alpha(C_1')$. Now
$\alpha$ and $\alpha\beta^{-1}$ are two new generators of the
Schottky group. The new fundamental domain is the region outside of
$C_1'$, $C_2'$, $\alpha(C_1')$ and $\alpha\beta^{-1}(C_2')$. The
quotient space is of course still the same genus two Riemann
surface, but its periods are acted by $A_1\to A_1+A_2$, $A_2\to
A_2$. Thus we see that a general $SL(2,{\bf Z})\subset
\Gamma_\infty$ corresponds to suitable redefinitions of the
generators of the Schottky group, but will leave ${\cal F}(\Omega)$
invariant.

Consider $f_{w,n}(\tau)$ a weight $w$ modular form, defined by
summing over modular images of $q^n$ (Poincar\'e series),
\eqn\modimag{ f_{w,n}(\tau) = P_w(e^{2\pi in\tau})= \sum_{\gamma\in
SL(2,{\bf Z})/\Gamma_\infty} (c\tau+d)^{-w} \exp\left(2\pi i n
{a\tau+b\over c\tau+d} \right) } We will first assume $4\leq w<12$,
so that the series converges absolutely and that there is no
elliptic cusp form weight $w$. In the limit $\tau\to i\infty$, the
only term in the sum that could remain finite is the term with
$\gamma={\bf 1}$, namely $q^n$. If $n>0$, this term vanishes in the
$\tau\to i\infty$ limit as well, we conclude that
$f_{w,n>0}(\tau)=0$. If $n\leq 0$, $f_{w,n}(\tau)$ has a
$q$-expansion of the form $q^n+{\cal O}(q)$. This uniquely fixes the
modular form $f_{w,n}(\tau)$, and its polar term is $q^n$. In the
cases $w=0,2$, the sum in \modimag\ can still make sense after
suitable regularization. Alternatively, one can always apply $D_w =
{1\over 2\pi i}\partial_\tau-{w\over 12}E_2(\tau)$ on a weight $w$
modular form to obtain a weight $w+2$ modular form, and bring a form
of weight $w\leq 2$ into the range $4\leq w<12$.